# Geometric Amplification via Non-Hermitian Berry Phase

J. R. Lane[1,*], C. Guria[1,*], J. Höller[2], T.D. Montalvo[1], Y. S. S. Patil[1], J. G. E. Harris[1,3,4,†]
[1] *Department of Physics, Yale University, New Haven, CT, USA*
[2] *Howard Hughes Medical Institute, Janelia Research Campus, Ashburn, VA, USA*
[3] *Department of Applied Physics, Yale University, New Haven, CT, USA*
[4] *Yale Quantum Institute, Yale University, New Haven, CT, USA*

Despite their apparent simplicity, coupled oscillators exhibit surprisingly complex phenomena. Two notable examples are Berry phase (a geometric or topological aspect of the oscillators' memory) and non-Hermiticity (the often counterintuitive impact of dissipation), both of which possess rich mathematical structures. Here, we demonstrate that combining Berry phase and non-Hermiticity leads to a fundamentally new form of amplification. Specifically, we show that this combination allows a lossy oscillator system to be converted into one with gain via slow modulation of its parameters. This is distinct from other amplification mechanisms, as it results specifically from the complex-valued Berry phase that is unique to non-Hermitian systems. We show that this mechanism produces continuous, useful gain in an optomechanical system, and that similar results can be realized in a very wide range of settings.

## I. INTRODUCTION

Geometric phase is a fundamental feature of classical and quantum dynamics [1,2,3,4,5,6,7,8,9] and plays an important role in photonics [1,10,11], crystalline materials [12,13,14,15], quantum information [16,17], and many-body physics [18,19,20]. Qualitatively, it can be viewed as an oscillating system's "memory" of how its parameters have been varied. More precisely, geometric phase is defined when: (*i*) the system's parameters are varied smoothly around a closed path (a loop) $\mathcal{C}$ in the space of these parameters, and (*ii*) at the end of this variation the system's state has changed by an overall phase factor $e^{-i\phi}$. Under these conditions, $\phi$ includes a contribution $\phi_{\mathrm{B}}$ (known as the geometric or Berry phase [8]) determined solely by the shape of $\mathcal{C}$ (i.e., it is independent of the duration of the traversal and the manner in which $\mathcal{C}$ is traversed). This robust relationship – between the shape of a loop and the effect of transporting a system around it – arises in many physical settings [7].

These aspects of geometric phase are independent of whether or not the system includes loss. However, in general there are dramatic differences between Hermitian systems (which do not include loss) and their non-Hermitian counterparts (which do) [21,22,23]. Notable examples include the topology of their spectra [24,25,26], their response to perturbations [27,28,29,30,31], and their performance in sensing and control applications [32,33,34].

Non-Hermiticity is also expected to alter the geometric phase, with perhaps the most striking difference being that $\phi_{\mathrm{B}}$ is a real number for Hermitian systems, while for non-Hermitian systems it is complex-valued [35]. This is an important distinction, as the imaginary part of $\phi_{\mathrm{B}}$ determines the amplitude of the oscillators' motion (while its real part only influences their phase). Thus, in addition to its interest as a fundamental aspect of coupled oscillators, accessing non-Hermitian $\phi_{\mathrm{B}}$ could result in novel geometric schemes for controlling the flow of energy in oscillator systems.

To date, most experiments on non-Hermitian systems have focused on static properties [24,25,26,36,37] or on non-geometric dynamics of the system's energy [38,39,40]. In contrast, phase evolution has been measured only in systems with access to a limited range of $\mathcal{C}$, precluding the observation of many non-Hermitian features of $\phi_{\mathrm{B}}$ [41,42,43]. Here, we present measurements of the complex-valued geometric phase in a non-Hermitian system with access to arbitrary $\mathcal{C}$. These measure-ments demonstrate a number of uniquely non-Hermitian features, including (gauge-invariant) geometric gain for $\mathcal{C}$ that are open paths (i.e., non-loops) [45], and the existence of an adiabatic limit for only one eigenmode [46,47].

* These authors contributed equally to this work.
† Contact author: jack.harris@yale.edu

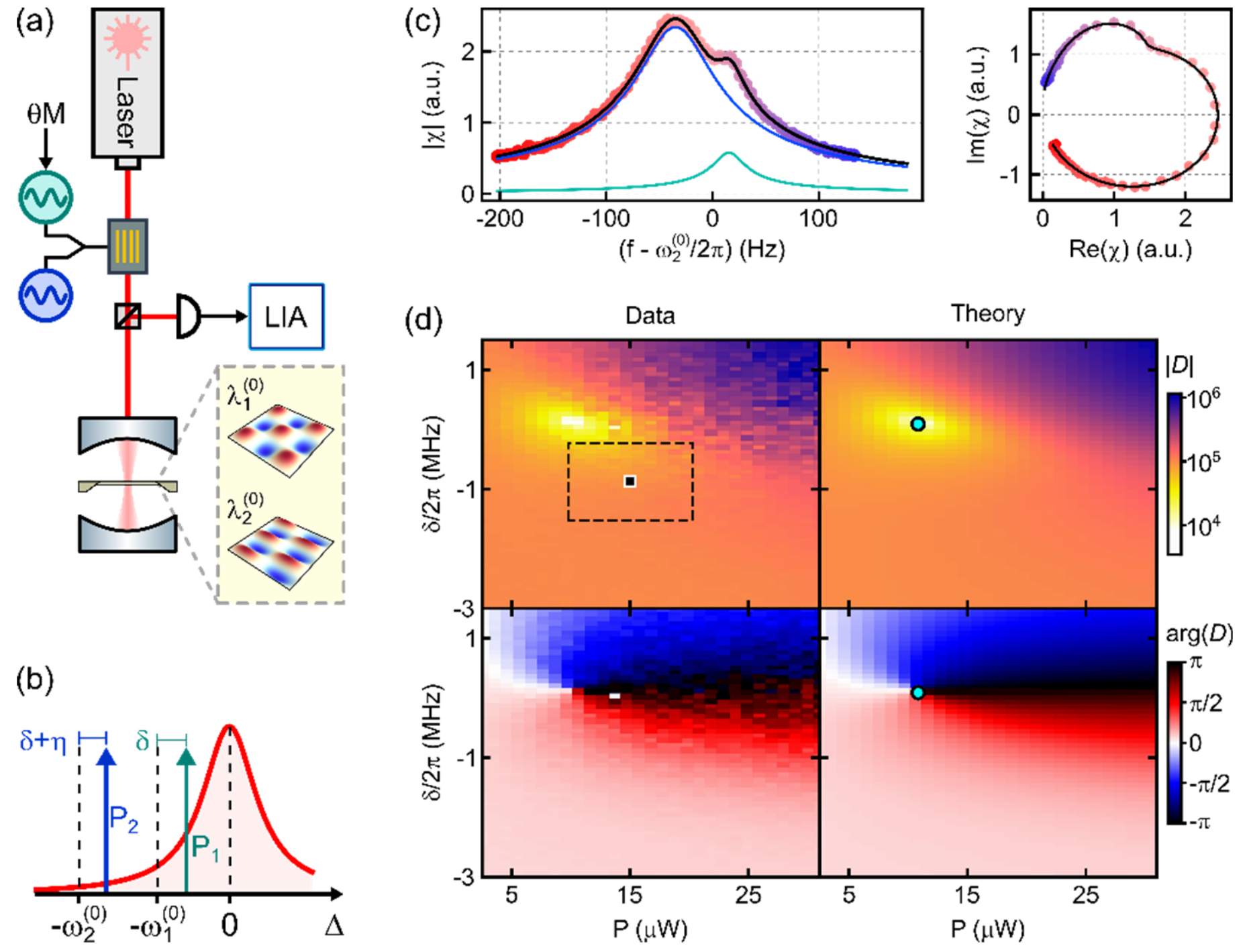

FIG. 1. Measurement setup. (a) Gray: $Si_3N_4$ membrane and cavity mirrors. Yellow: two vibrational modes of the membrane. Dark grey: acousto-optic modulator. Blue & green: rf sources. θM: phase modulation. PD: photodiode. LIA: lock-in amplifier. (b) Optical spectrum. Blue & green: control tones. Red: cavity mode. $P_i$: power of $i^{\text{th}}$ tone. $\delta$: the tones' common detuning. $\eta$: detuning offset. (c) Mechanical susceptibility $\chi$ vs. frequency $f$ for $P_{1,2} = P = 21.8$ μW, $\delta/2\pi = -1.2$ MHz. Left: $|\chi(f)|$ data (circles) and fit (black curve). Blue & green: the magnitude of each mode's contribution. Right: Parametric plot of $\chi(f)$ (circles: data, black curve: fit). Each data point's color indicates $f$. (d) $D(P, \delta)$. Left column: data. Right column: fit. Cyan circle: predicted exceptional point. Top row: $|D|$ (units: $(2\pi \times \text{Hz})^2$), bottom row: $\arg(D)$ (units: radians). Dashed box: the region corresponding to Fig. 3c. Black square: parameters used in Supplemental Material Fig. 10 [44]. In (c) and (d), $\eta/2\pi = -50$ Hz.

We also demonstrate a remarkable new feature of non-Hermitian geometric phase: it allows lossy linear elements to produce useful, continuous gain via the slow modulation of their parameters. This mechanism is fundamentally distinct from other forms of amplification, as it results specifically from the presence of loss in the system and from the geometric character of the dynamics. We show that this novel form of amplification does not require fine-tuning, and in fact is generic to non-Hermitian systems.

## II. RESULTS

### A. Experimental setup

The oscillators studied here are two vibrational modes of a $Si_3N_4$ membrane (500 μm × 500 μm × 150 nm). The membrane is inside an optical cavity (linewidth $\kappa/2\pi$ = 2.32 MHz) that is driven by a laser with wavelength $\lambda_{\text{opt}} = 1549.9$ nm (Fig. 1a). The two vibrational modes are tuned via their optomechanical interaction with the cavity (Supplemental Material Sec. 1 [44]; also Ref. [48]). The modes' bare eigenvalues (i.e., in the absence of any optomechanical effects) are $\lambda_1^{(0)}/2\pi = (2{,}423{,}969 - 1.8i)$ Hz and $\lambda_2^{(0)}/2\pi = (3{,}076{,}488 - 8.1i)$ Hz. Other device parameters are in Supplemental Material Sec. 2 and Table 1 [44].

The cavity is driven by two laser tones (Fig. 1b)

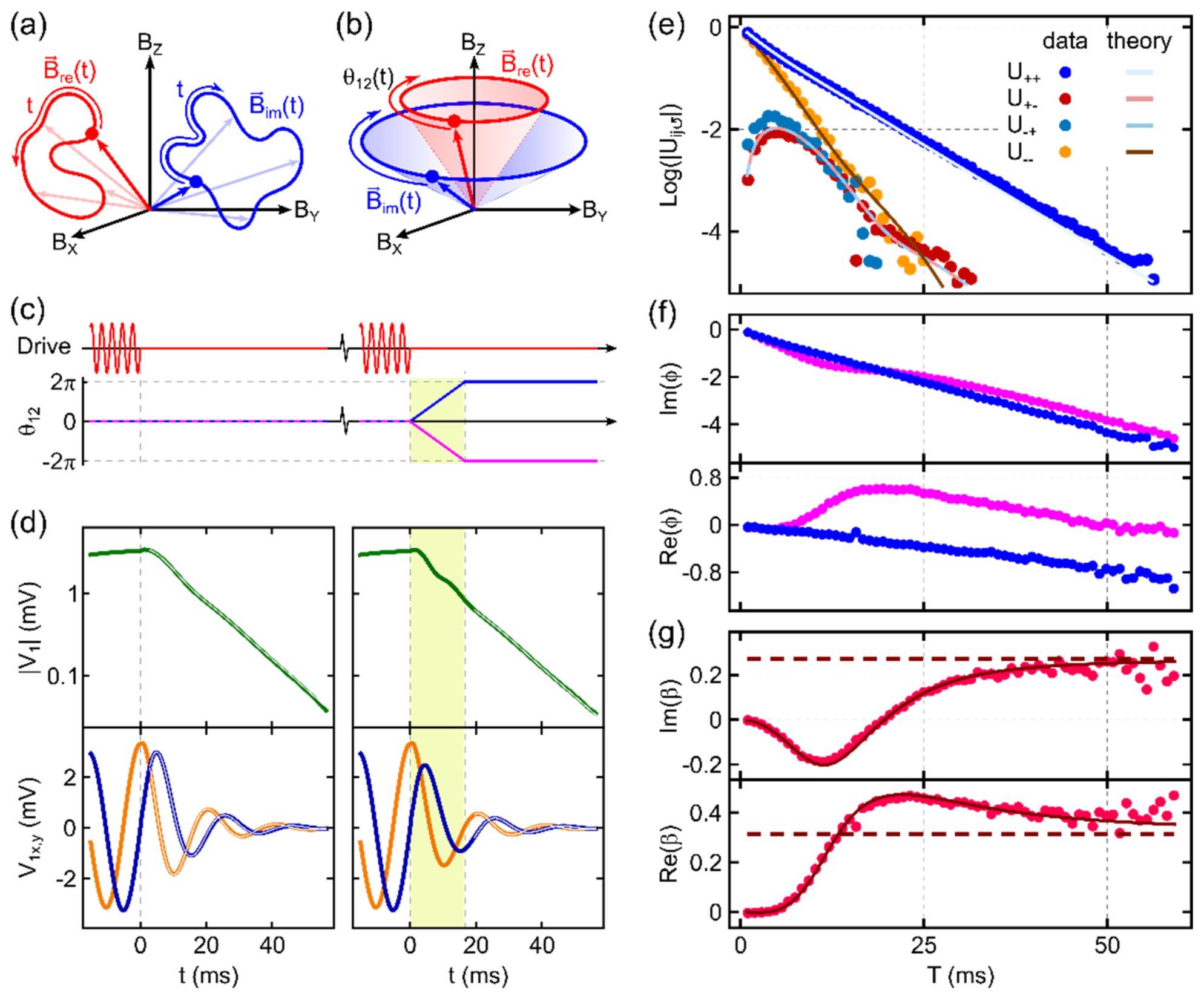

FIG. 2. Control loops and dynamics. (a) A generic control loop. Circles: the start and stop of the loop (at $t = 0, T$). Curves: $\vec{B}_{\mathrm{re,im}}(t)$. Arrows are a guide to the eye. (b) A "simple" control loop. (c) The timing of the experiment. Top: the drive that initializes the membrane's state. Bottom: $\theta_{12}(t)$ for $\mathcal{C}_{\circlearrowleft}$ (blue) and $\mathcal{C}_{\circlearrowright}$ (magenta). Yellow: the loop duration ($0 < t < T$). (d) The signal $V_1(t)$ from the membrane's motion. Here $T = 16.7$ ms. White lines: fit to the evolution at constant $H$ (Supplemental Material Sec. 3 [44]). $V_1$ reaches the noise floor for $t > 70$ ms. (e) The propagator matrix elements versus the loop duration $T$. (f) The complex phase $\phi(T)$ for $\mathcal{C}_{\circlearrowleft}$ (blue) and $\mathcal{C}_{\circlearrowright}$ (magenta). (g) $\beta(T)$. Solid line: theory. Dashed line: predicted $\phi_{\mathrm{B}}$ for $\mathcal{C}_{\circlearrowleft}$ (both without fit parameters). For (d-g), $P_{1,2} = P = 15$ μW, $\delta/2\pi = -1.0$ MHz, $\eta/2\pi = -50$ Hz.

with powers $P_{1,2}$ and detunings $\Delta_1 = -\omega_1^{(0)} + \delta$ and $\Delta_2 = -\omega_2^{(0)} + \delta + \eta$ where $\omega_{1,2}^{(0)} = \mathrm{Re}\left(\lambda_{1,2}^{(0)}\right)$. A small value of $|\eta|$ is used ($< 2\pi \times 100$ Hz) so that the frequency of the tones' beatnote $|\Delta_2 - \Delta_1| \approx |\omega_2^{(0)} - \omega_1^{(0)}|$. Qualitatively, each laser tone varies the mechanical modes' stiffness and damping, while the tones' beatnote provides tunable coupling between the modes [49,26].

In the standard description of cavity optomechanics [48] the time evolution of the two mechanical modes obeys:

$$\dot{\vec{c}} = -iH\vec{c} \tag{1}$$

Here $\vec{c}(t) = \left(c_1(t), c_2(t)\right)^T$ and $c_{1,2}$ are the modes' amplitudes. The non-Hermitian $2 \times 2$ matrix $H$ is tuned via the control parameters $\vec{X} \equiv (P_1, P_2, \delta, \eta, \theta_{12})$ where $\theta_{12}$ is the phase of the intracavity beatnote. In a frame that rotates with this beatnote, $H$ depends on time only via $\vec{X}(t)$ (Supplemental Material Sec. 1 [44]).

The membrane's motion is measured via standard heterodyne techniques that produce two signals $V_{1,2}(t) \propto c_{1,2}(t)$ (Appendix). The mechanical modes are characterized via their response to an oscillating force (Fig. 1c), which is fit to determine $\lambda_\pm$ (the eigenvalues of $H$). Similar measurements of $\lambda_\pm$ over a range of $\delta$ and $P_1 = P_2 = P$ are summarized in Fig. 1d, which shows the discriminant $D = (\lambda_+ - \lambda_-)^2$. Fitting this to $D(\delta, P)$ calculated from $H(\vec{X})$ (Supplemental Material Secs. 1,2 [44]) returns the parameter values in Supplemental Material Table 1 [44].

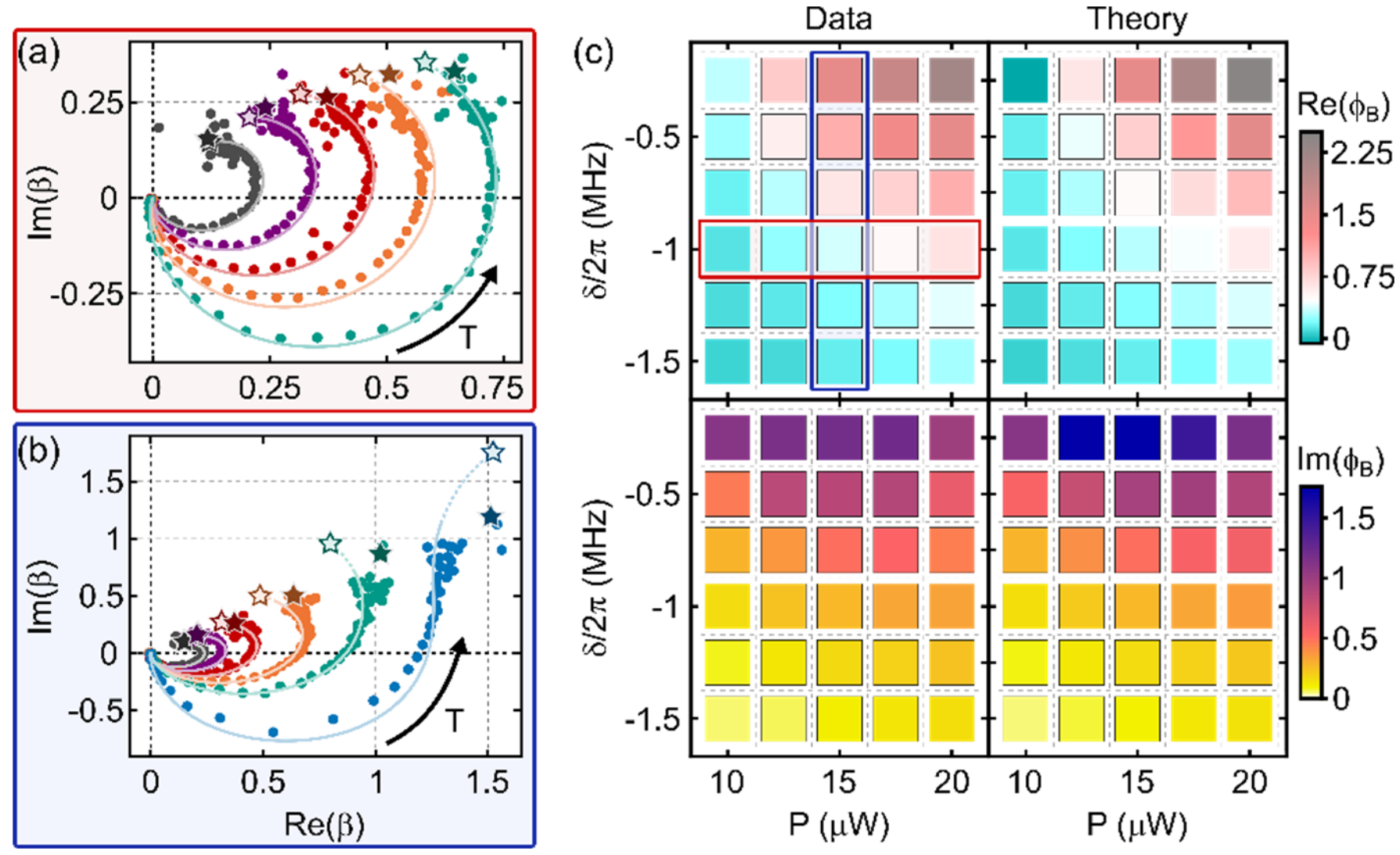

FIG. 3. Measuring the complex geometric phase. (a) Parametric plots of $\beta(T)$. Data nearest the origin corresponds to $T = 0$. Here $\delta/2\pi = -1$ MHz and $P_{1,2} = P = (10,12.5,15,17.5,20)$ μW (gray, purple, red, orange, teal). Solid line: predicted $\beta(T)$. Hollow star: predicted $\phi_{\mathrm{B}}$. Solid star: measured $\phi_{\mathrm{B}}$. (b) As in (a), but $\delta/2\pi = (-1.5, -1.25, -1, -0.75, -0.5, -0.25)$ MHz (gray, purple, red, orange, teal, blue) and $P_{1,2} = P = 15$ μW. (c) The value of $\phi_{\mathrm{B}}$ from measurements as in (a,b) for a range of $(P, \delta)$. Data in the red (blue) box corresponds to measurements in (a) ([b]). For all measurements, $\eta/2\pi = -50$ Hz.

### B. Visualizing non-Hermitian control loops

In a Hermitian two-mode system, $H$ can be represented (ignoring its trace) as a three-vector $\vec{B}$ defined via $H = \vec{B} \cdot \vec{\sigma}$ (where $\vec{\sigma}$ is the Pauli vector) , and a control loop simply corresponds to the path in $\mathbb{R}^3$ taken by $\vec{B}$. This picture can be extended to non-Hermitian systems by representing $H$ as a pair of three-vectors: $H = \left(\vec{B}_{\mathrm{re}} + i\vec{B}_{\mathrm{im}}\right) \cdot \vec{\sigma}$. In this case, a control loop corresponds to $\vec{B}_{\mathrm{re,im}}$ each tracing out a loop in $\mathbb{R}^3$ (Fig. 2a).

In our setup $\vec{B}_{\mathrm{re,im}}$ are set by $\vec{X}$, and a loop $\mathcal{C}$ is defined parametrically via $\vec{X}(s)$, where $0 \le s \le 1$ and $\vec{X}(0) = \vec{X}(1)$. A traversal of $\mathcal{C}$ is given by choosing a duration $T$ and a function $s(t/T)$, with $s(0) = 0$ and $s(1) = 1$. A particularly simple class of loops consists of holding $P_1, P_2, \delta, \eta$ constant while $\theta_{12} = 2\pi s$ and $s = t/T$ (this is a loop because $\theta_{12}$ is defined mod $2\pi$). These "simple" loops consist of $\vec{B}_{\mathrm{re,im}}$ rotating at a constant rate once around the $z$ axis (Fig. 2b).

### C. Measuring the geometric phase

To measure the geometric phase associated with a loop, we determine the propagator matrix $U(T)$ that fully characterizes the modes' evolution during the loop. First, the membrane is driven at frequency $\omega_{\mathrm{d}}^{(a)} \approx \omega_1^{(0)}$ (or $\omega_2^{(0)}$) until it reaches a steady state $\vec{c}^{(a)}(0)$. The drive is then turned off at $t = 0$ and the membrane's subsequent motion is recorded and fit to determine $\vec{c}^{(a)}(0)$ (Supplemental Material Sec. 3 [44]). Then the membrane is re-initialized the same way and the drive is turned off (again at $t = 0$), and a time-dependent phase shift is applied to one of the laser tones so that $\theta_{12}(t) = 2\pi t/T$, thereby meeting condition ($i$). After this loop is completed (i.e., for $t > T$) $\theta_{12}$ is held constant at $2\pi$, and the membrane's motion for $t > T$ is fit to determine $\vec{c}^{(a)}(T)$. This sequence is shown in Fig. 2c, with the corresponding data in Fig. 2d.

$U(T)$ is determined by repeating these steps with slightly different drive frequency ($\omega_{\mathrm{d}}^{(b)}$) to produce different initial and final states $\vec{c}^{(b)}(0,T)$ (Supplemental Material Sec. 3 [44]). Fig. 2e shows $U(T)$ measured in this way, with each color representing an element of $U$ in the basis of $\vec{\psi}_{\pm}$ (the less-damped and more-damped eigenmodes of $H(0)$, respectively). The system's overall damping is evident in the decay of all four elements with increasing $T$.

For large $T$, $U_{++}$ dominates the other matrix elements, reflecting the fact that in a non-Hermitian system the adiabatic theorem applies only to a state if it is the least-damped eigenmode of $H$ throughout $\mathcal{C}$ [46,47]. A given loop possesses such a state if it does not cross a "gain/loss boundary" (i.e., at which $\mathrm{Im}(\lambda_+) = \mathrm{Im}(\lambda_-)$); this condition is met by all the loops used in this study.

For evolution governed by Eq. 1, the complex phase $\phi(T)$ defined via $U_{++}(T) = e^{-i\phi(T)}$ is predicted to be (for large $T$):

$$\phi(T) = q_{\mathrm{D}} T - \phi_{\mathrm{B}} + q_1 T^{-1} + q_2 T^{-2} + \cdots \qquad (2)$$

where the $q_i$ are complex constants (Supplemental Material Sec. 4 [44]; also Refs. [46,47]). We refer to the first term $q_{\mathrm{D}} T = \phi_{\mathrm{D}}$ as the dynamical phase. For a simple loop $q_{\mathrm{D}} = \lambda_+$ (the eigenvalue associated with $\vec{\psi}_+$), while for a non-simple loop $q_{\mathrm{D}}$ is the time-average of $\lambda_+$. In practice, Eq. 2 is expected to hold when the adiabaticity condition $T \gtrsim T_{\mathrm{ad}}$ is satisfied, where $T_{\mathrm{ad}}$ is the largest value of $|\lambda_+(s) - \lambda_-(s)|^{-1}$ along $\mathcal{C}$ [46,50].

Figure 2f shows $\phi(T)$ as determined from $U_{++}(T)$. At large $T$, the amplitude of the motion decays roughly exponentially with $T$ (upper panel) while its phase evolves roughly linearly (lower panel), reflecting the fact that $\phi_{\mathrm{D}}$ dominates $\phi(T)$ for large $T$ (Eq. 2).

To isolate $\phi_{\mathrm{B}}$, we also perform measurements that are identical except that $\theta_{12}(t) = -2\pi t/T$. Loosely speaking, this amounts to traversing the same loop with the opposite sense; however, these loops are formally inequivalent, and we denote them as $\mathcal{C}_{\circlearrowleft,\circlearrowright}$ (for $\dot{\theta}_{12} \gtrless 0$). For two such loops, the odd-order terms in Eq. 2 are identical (e.g., $q_{\mathrm{D},\circlearrowleft} = q_{\mathrm{D},\circlearrowright}$ and $q_{1,\circlearrowleft} = q_{1,\circlearrowright}$) while the even-order terms differ by a sign (e.g., $\phi_{\mathrm{B},\circlearrowleft} = -\phi_{\mathrm{B},\circlearrowright}$ and $q_{2,\circlearrowleft} = -q_{2,\circlearrowright}$) (Supplemental Material Sec. 4 [44]).

Consistent with this, the measured $\phi_{\circlearrowleft}(T)$ and $\phi_{\circlearrowright}(T)$ show similar trends at large $T$, but are clearly offset from each other (Fig. 2f). Figure 2g shows $\beta(T) = \frac{1}{2}\big(\phi_{\circlearrowright}(T) - \phi_{\circlearrowleft}(T)\big)$, which should approach $\phi_{\mathrm{B},\circlearrowleft}$ (the geometric phase of $\mathcal{C}_{\circlearrowleft}$) for large $T$. These measurements agree well with the no-free-parameter predictions.

### D. Varying the control loop

To demonstrate the geometric character of $\phi_{\mathrm{B}}$, we alter $\mathcal{C}$ by performing the same real-time variation of $\theta_{12}(t) = \pm 2\pi t/T$ as in Fig. 2c, but for different $(P,\delta)$. This corresponds to loops as in Fig. 2b but with different $\vec{B}_{\mathrm{re,im}}(0)$. The resulting $\beta(T)$ are shown in Fig. 3a,b.

For each of these loops, $\phi_{\mathrm{B},\circlearrowleft}$ is determined by fitting $\beta(T)$ at large $T$ to the asymptotic form of Eq. 2: $\beta(T) = \phi_{\mathrm{B},\circlearrowleft} - q_{2,\circlearrowleft}/T^2$. The values of $\phi_{\mathrm{B},\circlearrowleft}$ (denoted as $\phi_{\mathrm{B}}$ from here on) returned by these fits are the solid stars in Fig. 3a,b. Similar measurements over a wider range of $(P,\delta)$ are shown in Fig. 3c (left-hand panels), along with the no-free-parameter predictions (right-hand panels) calculated using the parameters in Supplemental Material Table 1 [44].

The geometric character of $\phi_{\mathrm{B}}$ is also evident in measurements that use more complicated loops. This includes $\mathcal{C}$ in which $P_2, \delta,$ and $\theta_{12}$ are varied simultaneously (i.e., "non-simple" loops as in Fig. 2a) and $\mathcal{C}$ with non-linear $\theta_{12}(t)$ (Supplemental Material Sec. 6 [44]). It also includes control paths that are not closed loops (Supplemental Material Sec. 7 [44]), for which $\mathrm{Im}(\phi_{\mathrm{B}})$ has a gauge-independent definition (even though $\mathrm{Re}(\phi_{\mathrm{B}})$ does not) [45]. All of these measurements show good agreement with the no-free-parameters predictions.

### E. Steady-state geometric gain

The imaginary part of the geometric phase directly impacts the amplitude of the normal mode's motion. This effect has no analog in Hermitian systems, and so represents a novel means for controlling the oscillators' energy. In particular, the geometric phase contributes gain to the normal mode for any loop with $\mathrm{Im}(\phi_{\mathrm{B}}) < 0$. In the measurements described so far, this contribution is overwhelmed by the mode's intrinsic damping. For example, in Fig. 2f the amplitude of motion always decreases during $\mathcal{C}_{\circlearrowright}$ even though $\mathrm{Im}\big(\phi_{\mathrm{B},\circlearrowright}\big) < 0$.

To produce net amplification of the motion, the loop's geometric gain should exceed the dynamical loss: $0 < -\mathrm{Im}(\phi_{\mathrm{D}}) < -\mathrm{Im}(\phi_{\mathrm{B}})$. This condition cannot be met in the large-$T$ limit with a given $\mathcal{C}$ because $\mathrm{Im}(\phi_{\mathrm{B}}) \propto T^0$ while $\mathrm{Im}(\phi_{\mathrm{D}}) \propto T^1$ (Eq. 2). Nevertheless, we show here that it is possible to construct $\mathcal{C}$ such that the geometric gain amplifies the normal mode's motion for arbitrarily large $T$.

To demonstrate this, we use the same measurement sequence (Fig. 2c) and setup (Fig 1a), except that the $\lambda_2^{(0)}$ mode is replaced by a mode with bare eigenvalue $\lambda_3^{(0)}/2\pi = (3{,}331{,}064 - 1.92i)$ Hz (see Supplemental Material Table 1 [44] for its other parameters). In addition, a third laser tone is added (Fig. 4a) to access a greater range of $\vec{B}_{\mathrm{re,im}}(t)$. The system is transported around the loop $\mathcal{C}_{\mathrm{amp}}$ shown in Fig. 4b,c by varying $P_2$ and $\theta_{12}$ while holding the other parameters constant.

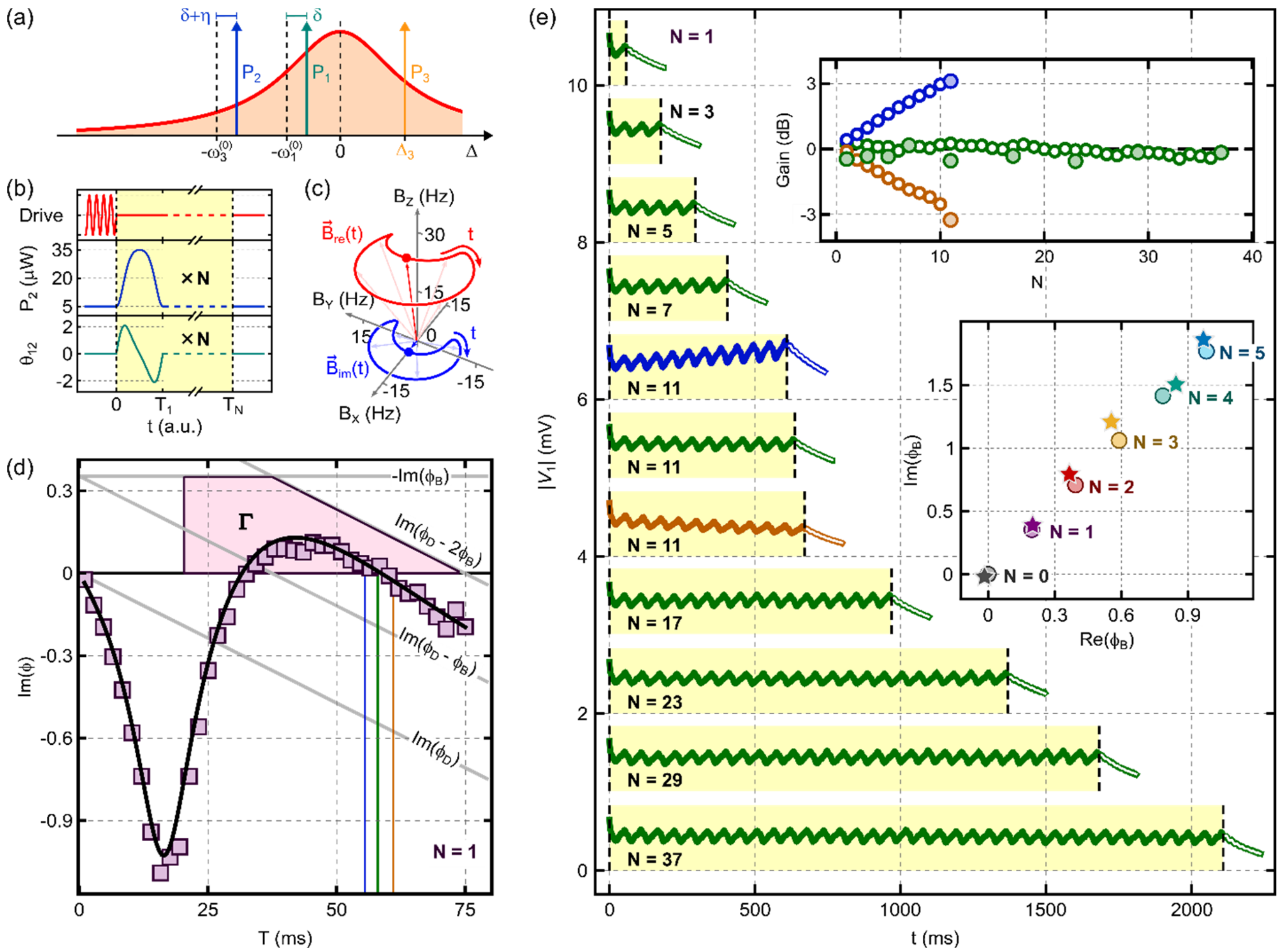

FIG. 4. Steady-state geometric gain. (a) The control tones. (b) $P_2(t)$ and $\theta_{12}(t)$ during a loop $\mathcal{C}_{\mathrm{amp}}^{(N)}$. (c) The loop $\mathcal{C}_{\mathrm{amp}}^{(1)}$ visualized as $\vec{B}_{\mathrm{re,im}}(t)$. (d) The gain Im[ϕ] when $\mathcal{C}_{\mathrm{amp}}^{(1)}$ is traversed in time $T$. Squares: data. Black curve: theory. Pink: the region Γ that produces SSGG (Supplemental Material Sec. 8 [44]). Blue, green, orange lines: values of $T$ used in (e). (e) The amplitude of motion during loops $\mathcal{C}_{\mathrm{amp}}^{(N)}$ with $T_1 = (55.5, 58, 61)$ ms (blue, green, orange). Curves are offset by 1 mV, so $|V_1| = 0$ is at the bottom of each yellow band. Black dashed lines: the duration $T$ of the control sequence. White curves: fits to the evolution at constant $H$ after $\mathcal{C}_{\mathrm{amp}}^{(N)}$ ends. Upper inset: the gain resulting from these loops. Filled circles: the gain from each $\mathcal{C}_{\mathrm{amp}}^{(N)}$ in the main panel. Green hollow circles: the gain after each circuit of the $N = 37$ loop. Blue & orange hollow circles: the same for the $N = 11$ loop with $T_1 = (55.5, 61)$ ms (blue, orange) For (d,e), $P_1 = 26\ \mu$W, $P_3 = 28\ \mu$W, $\Delta_3/2\pi = 3.38$ MHz, $\delta/2\pi = -0.911$ MHz, $\eta/2\pi = -50$ Hz. Lower inset: predicted $\phi_{\mathrm{B}}$ (circles); measured $\phi_{\mathrm{B}}$ (stars) for $\mathcal{C}_{\mathrm{amp}}^{(N)}$.

Figure 4d shows the gain produced by traversing $\mathcal{C}_{\mathrm{amp}}$. Consistent with Eq. 2, the dynamical loss dominates at large $T$, leading to negative gain for $T >$ 58 ms. However, there is a range of $T$ over which the mode experiences positive gain (32 ms $< T <$ 58 ms) despite the loss contributed by $\mathrm{Im}(\phi_{\mathrm{D}})$.

This positive gain can be extended to arbitrarily large times (i.e., to the steady state) simply by repeating the loop. This is demonstrated in Fig. 4e, which shows the mode's amplitude of motion while $\mathcal{C}_{\text{amp}}$ is repeated $N$ times, with each repetition occurring in time $T_1$ for a total duration $T = NT_1$.

The green data show measurements for $T_1 = 58$ ms and with $N$ as large as 37 (corresponding to $T > 2$ s). In this data, the amplitude of motion returns to the same value after every repetition of $\mathcal{C}_{\text{amp}}$. For similar measurements using $T_1 = 55.5$ ms (blue data) the mode's motion grows each time $\mathcal{C}_{\text{amp}}$ is repeated, while for $T_1 = 61$ ms (blue data) the mode's motion decreases. This reflects the fact that $T_1 = 58$ ms corresponds to the "break-even" between the gain and the dynamical loss produced by a traversal (Fig. 4d), while slightly smaller (larger) values of $T_1$ result in net gain (loss) for each traversal.

The complex geometric phase measured for $\mathcal{C}_{\text{amp}}^{(N)}$ is shown in the lower inset of Fig. 4e. The data agree well with the calculated value of $N\phi_{\text{B,amp}}$ (where $\phi_{\text{B,amp}}$ is the predicted geometric phase for $\mathcal{C}_{\text{amp}}$), indicating that the mode's oscillations are phase-coherent during this control sequence.

## III. DISCUSSION

The measurements shown in Fig. 4e represent the geometric phase dominating over the dynamical phase in the long-time limit. This may seem to contradict Eq. 2, which predicts the opposite. However, Eq. 2 applies to a given loop, while each of the green sequences in Fig. 4e (i.e., each choice of $N$) formally corresponds to a distinct loop. If we denote these loops by $\mathcal{C}_{\text{amp}}^{(N)}$ (with $\mathcal{C}_{\text{amp}}^{(1)} = \mathcal{C}_{\text{amp}}$), then gain results from traversing $\mathcal{C}_{\text{amp}}^{(N)}$ in time $T = NT_1$, with $T_1$ chosen so that $\mathcal{C}_{\text{amp}}^{(1)}$ has net gain (i.e., $\text{Im}[\phi(T_1)] > 0$) and with no limit on how large $N$ (and hence $T$) may be.

The results in Fig. 4e can also be understood by noting that the $N \to \infty$ limit of the sequence in Fig. 4b corresponds to taking $T \to \infty$ while fixing the parameter variation rate $\dot{\vec{B}}_{\text{re,im}}$ (the total variation remains finite because the control sequence is cyclic). This is not the standard adiabatic limit (in which $T \to \infty$ with fixed $T\dot{\vec{B}}_{\text{re,im}}$), but the adiabatic approximation is still accurate (and hence $\phi_{\text{B}}$ is relevant) as long as $\dot{\vec{B}}_{\text{re,im}}$ is sufficently small.

This is the key distinction between the steady-state geometric gain (SSGG) demonstrated here and previous studies of gain produced by $\text{Im}(\phi_{\text{B}})$, both experimental [42] and theoretical [51,52]. These worked in the standard adiabatic limit, in which a very long time ($T \to \infty$) is used to produce a fixed amount of gain. In this case, any non-zero dynamical loss results in net damping. Equivalently, gain occurs only if the system has been fine-tuned to have lossless eigenmodes [42,51,52].

In contrast, Fig. 4e shows that SSGG overcomes finite damping and does not require fine tuning. This can be seen qualitatively by comparing the geometric gain of a generic loop with the dynamical loss incurred by traversing the loop slowly enough for the adiabatic approximation to be accurate. Taking the former to be $\sim 1$ (as in Fig. 3c) and the latter to be $\sim \gamma_+ T_{\text{ad}}$ (where $\gamma_\pm = -\text{Im}(\lambda_\pm)$ are the eigenmodes' damping rates along the loop), net gain occurs when $\gamma_+ T_{\text{ad}} \lesssim 1$. Since $T_{\text{ad}} \lesssim \delta\gamma^{-1}$ (where $\delta\gamma \equiv \gamma_+ - \gamma_-$), SSGG can occur for loops in which $2\gamma_+ \lesssim \gamma_-$. A quantitative analysis agrees with this picture, and shows that a wide range of loops, both simple and non-simple, produce SSGG (Supplemental Material Sec. 8 and Fig. 5 [44]).

This picture highlights a number of important differences between SSGG and other mechanisms that add gain to an oscillator via modulation of the system's parameters. For example, conventional parametric amplification [53] (PA) uses a modulation rate $\sim 2\omega_0$, which is larger than the typical rate required for SSGG by a factor $\sim Q$ (the mode's quality factor). PA amplifies only one quadrature of a signal, while SSGG is phase-insensitive. Lastly, these two amplification mechanisms are fundamentally distinct, as PA does not involve adiabaticity, geometric phase, or non-Hermiticity.

Finally, we note that SSGG occurs only because the geometric phase is complex, which in turn requires the presence of loss in the system (more precisely, it requires $\delta\gamma > 0$). Thus, SSGG can be viewed as the direct conversion of a system's loss into useful, continuous gain via the slow modulation of its parameters.

## ACKNOWLEDGMENTS

We thank C. D. Brown, V. Chavva, S. M. Girvin, J. Kim, K. W. Lehnert, K. Murch, A. Neely, A. J. Pillai, N. Read, H. Ribeiro, C. C. Wilson, and Y. Zhang for discussions. This work is supported by Air Force Office of Scientific Research (FA9550-21-1-0152), the Vannevar Bush Faculty Fellowship (N00014-20-1-2628), Air Force Office of Scientific Research Multidisciplinary University Research Initiative (FA9550-21-1-0202), and the Yale University Mossman Prize Fellowship in Physics.

## DATA AVAILABILITY

The data are not publicly available. The data are available from the authors upon reasonable request.

## APPENDIX: EXPERIMENTAL SETUP

**1. The cavity optomechanical system.** The device consists of a 500 μm × 500 μm × 150 nm silicon nitride membrane inside an optical Fabry-Pérot cavity. The membrane is mounted near to the cavity's waist, and with the cavity's axis normal to the membrane.

The cavity is formed by two mirrors separated by 7 mm, each with radius of curvature 50 mm. The mirrors' specified reflectivity $R = 0.99975$ and loss $\sim$5 ppm. With the membrane inside the cavity, the finesse $\mathcal{F} = 10{,}500$.

The mirrors are mounted on piezo elements, which are used to ensure that the membrane is located roughly midway between a node and an antinode of the cavity's optical mode (to maximize the linear coupling between the cavity mode and the membrane's motion). The laser drives a $\mathrm{TEM}_{00}$ mode of the cavity, with input coupling rate $\kappa_{\mathrm{in}}/\kappa = 0.346$.

The experiment is housed in a room-temperature vacuum chamber with pressure $P < 6 \times 10^{-8}$ mBar. To minimize drifts in the device's parameters, it is mounted on a thermoelectric cooler that stabilizes the device's temperature to within 1 mK.

The membrane's normal vibrational modes are indexed as $(m, n)$ where $m$ and $n$ are the number of antinodes in the $x$ and $y$ directions, respectively. We use the $(3,3)$, $(5,2)$ and $(5,3)$ modes, which we refer to as "mode 1", "mode 2", and "mode 3". Their bare resonant frequencies $\omega_1^{(0)}, \omega_2^{(0)}, \omega_3^{(0)}$ and bare decay rates $\gamma_1^{(0)}, \gamma_2^{(0)}, \gamma_3^{(0)}$ are given in Supplemental Material Table 1 [44].

For all the measurements described here, both the optomechanically-induced shift in the mechanical modes' resonance frequencies and the optomechanically-induced damping rate are $\lesssim 2\pi \times 10^2$ Hz.

**2. Laser control and readout.** All of the laser tones are generated from an NKT Adjustik laser, which is detuned by $-83$ MHz from the cavity mode of interest (Supplemental Material Fig. 1 [44]). The laser is split into several tones that are recombined before being sent to the cavity. The first tone ("probe") is shifted +83 MHz by an acousto-optic modulator (pAOM, Gooch and Housego FiberQ) driven by function generator FG1 (RIGOL DG4162). This tone is locked to the cavity resonance using standard Pound-Drever-Hall (PDH) techniques with phase modulation sidebands at ±32.5 MHz created by an electrooptic modulator (EOM, Thorlabs LN65S-FC) driven by a laser lock box (MOKU:Lab).

This tone is recombined with the unshifted laser, which serves as a local oscillator (LO) for the heterodyne measurements (described below) before being sent to one arm of a polarizing beam splitter (PBS, Thorlabs PBS PFC1550A). To prevent low-frequency power fluctuations, both the probe and LO beams are stabilized by monitoring each beam's power and using a PID controller (New Focus LB10005) to tune an attenuator (VOA) (Thorlabs V1550A and V1550PA for the probe and LO, respectively).

We also optimize the polarization of each beam with paddle tensioners (Thorlabs FPC032), and pass the combined LO and probe beam through an in-line polarizer (Thorlabs ILP1550PM-APC) to optimize the input to the polarizing beam splitter.

Another tone ("control") is power-stabilized and sent through a second AOM (cAOM, Gooch and Housego Fiber-Q) which is driven by two microwave tones at 83 MHz $- \omega_1^{(0)}/(2\pi) + \delta/(2\pi)$ and 83 MHz $- \omega_2^{(0)}/(2\pi) + \delta/(2\pi) - \eta/(2\pi)$ produced by FG2 and FG3 respectively (RIGOL DG4162). The amplitude and frequency of these microwave tones define the control parameters $(P_1, P_2, \delta, \eta)$. The tone from FG2 is phase modulated by an arbitrary waveform generator (AWG, RIGOL DG4162) to control $\theta_{12}$.

The control tone is again polarization optimized and sent to the other arm of the PBS to ensure that the probe and LO have orthogonal polarization from the control tone. This prevents the control tone from contaminating the heterodyne signal. It also prevents spurious driving of the membrane that would result from intracavity beating between the control and probe tones.

The output of this PBS is sent to a collimator (Thorlabs CFC2A-C) and is mode matched to a $\mathrm{TEM}_{00}$ mode of the optical cavity. The light from the cavity passes through a 50:50 beam splitter (Thorlabs BS015) and a polarizer (Thorlabs LPNIR050-MP2) to remove the control tone prior to being recorded by a photodetector (PD5, Thorlabs PDA10CF). The electrical signal from this photodetector is split, filtered, and sent to the lock box, where the signal at 32.5 MHz is mixed down to DC to generate a PDH error signal, which is fed to the laser's piezo driver.

When the membrane oscillates at a frequency ω, it imparts phase modulation sidebands on the probe. The beatnote between the probe and LO is detected by PD5 which converts it to an electrical signal at 83 MHz $\pm\,\omega/(2\pi)$. This is mixed with a tone at 99.354 MHz (FG4, Vaunix LMS-451D-13) and bandpass filtered to isolate the heterodyne signal at 16.354 MHz $\pm\,\omega/(2\pi)$. This is sent to the LIA, where it is mixed down by two demodulation oscillators: one at 16.354 MHz $+\,\omega_1^{\mathrm{mod}}/(2\pi)$ and one at 16.354 MHz$+\omega_2^{\mathrm{mod}}/(2\pi)$, with $\omega_1^{\mathrm{mod}} \approx \omega_1^{(0)}$ and $\omega_2^{\mathrm{mod}} \approx \omega_2^{(0)}$. This converts each quadrature of the beatnote at 16.354 MHz $+\,\omega/(2\pi)$ to a voltage signal from which we extract the complex motional amplitudes.

To drive the membrane's motion, we use a tone with frequency $\omega_{\mathrm{d}}/(2\pi) \sim \omega_i^{(0)}/(2\pi)$ generated by the lock-in amplifier (LIA, Zurich Instrument HF2) to amplitude modulate (AM) the 83 MHz tone that drives the pAOM. The AM tone is gated using a fast microwave switch (Minicircuits ZASWA-2-50DRA+) which is toggled by a TTL signal. The falling edge of this TTL signal triggers the start of the control loop.

Supplemental Material for

# Geometric amplification via non-Hermitian Berry phase

J.R. Lane[1]†, C. Guria[1]†, J. Höller[2], T.D. Montalvo[1], Y.S.S. Patil[1], J.G.E. Harris[1,3,4]*

[1] Department of Physics, Yale University, New Haven, CT, USA

[2] Howard Hughes Medical Institute, Janelia Research Campus, Ashburn, VA, USA

[3] Department of Applied Physics, Yale University, New Haven, CT, USA

[4] Yale Quantum Institute, Yale University, New Haven, CT, USA

* Corresponding author. Email: jack.harris@yale.edu

† These authors contributed equally to this work.

**Supplemental Note 1: Optomechanical model**

This Supplemental Note describes the theoretical model of the optomechanical system. This model includes the interactions between the membrane and the optical cavity, as well as the manner in which the laser tones appear as control parameters in the mechanical modes' dynamical matrix (which is commonly referred to as its "Hamiltonian").

### *1.1 The Hamiltonian*

We consider a system of two mechanical modes coupled to a single optical cavity mode. The classical Hamiltonian function for the full system is given by:

$$H_{\text{full}} = \hbar\left(\Omega - i\frac{\kappa}{2}\right)a^*a + \sum_{j=1}^{2}\hbar\left(\omega_j^{(0)} - i\frac{\gamma_j^{(0)}}{2}\right)\tilde{c}_j^*\tilde{c}_j + \sum_{j=1}^{2}\hbar g_j\left(\tilde{c}_j + \tilde{c}_j^*\right)a^*a + \sum_{j=1}^{2}\hbar\left(A_j a^* a\right)\tilde{c}_j^*\tilde{c}_j \tag{S1}$$

Where $\Omega$ is the resonance frequency of the optical mode, $a$ is the complex-valued amplitude of the optical mode driven by the control laser, $\tilde{c}_j$ are the complex-valued amplitudes of the two mechanical modes, and * denotes complex conjugation. The first two terms correspond to the uncoupled optical and mechanical oscillators respectively. The third term corresponds to radiation pressure coupling between the two with coupling strengths $g_j$. The last term (which we refer to as photothermal) describes the temperature-dependence of the mechanical modes' resonance frequencies and the heating of the membrane by the control laser. We note that the appearance of the reduced Plank's constant $\hbar$ serves only to conform with the broader literature on optomechanics, where the coupling strengths $g_j$ are usually given as single-photon rates. For consistency, the thermal response coefficients $\mathrm{A}_j$ are also given as single-photon rates. Values of these parameters (as determined by measurements described in Supplemental Note 2) are provided in Table S1.

The dynamics of $a$ and $c_j$ are governed by Hamilton's equations of motion. In particular, the optical mode $a$ is driven by two control laser tones (with input coupling $\kappa_{\text{in}}$) as given below:

$$a_{\text{in}} = e^{-i\Omega t}\sum_{n=1}^{2}\sqrt{\frac{\mathrm{P}_n}{\hbar(\Omega + \Delta_n)}}\,e^{-i(\Delta_n t + \theta_n)} \tag{S2}$$

where $P_n, \Delta_n, \theta_n$ are the power, detuning (from the cavity resonance) and phase of the $n^{\text{th}}$ control tone, respectively. For our system, $\kappa \gg (g_j, \gamma_j, A_j)$, so we can linearize the optomechanical coupling term and adiabatically eliminate the optical field.

Furthermore, the intensity beatnote between the two control tones is at frequency $|\Delta_{12}| =$

$|\Delta_1 - \Delta_2| \approx \left|\omega_1^{(0)} - \omega_2^{(0)}\right|$, which provides coupling between the two mechanical modes. Under the rotating wave approximation [1,2] the resulting effective equation of motion for the two mechanical modes is given by:

$$i\dot{\vec{c}}_{\text{lab}}(t) = H_{\text{lab}}(t)\vec{c}_{\text{lab}}(t) \tag{S3}$$

in the lab frame, where $\vec{c}_{\text{lab}}(t) = (\tilde{c}_1(t), \tilde{c}_2(t))^T$ denotes the mechanical modes' complex-valued amplitudes and

$$H_{\text{lab}}(t) = \begin{pmatrix} \omega_1^{(0)} - \frac{i\gamma_1^{(0)}}{2} + \sum_{n=1}^{2} \sigma_{nn,1}(t) + A_1 n_{\text{op}}(t) & \sigma_{12}(t)e^{i(\Delta_{12}(t)t+\theta_{12}(t))} \\ \sigma_{21}(t)e^{-i(\Delta_{12}(t)t+\theta_{12}(t))} & \omega_2^{(0)} - \frac{i\gamma_2^{(0)}}{2} + \sum_{n=1}^{2} \sigma_{nn,2}(t) + A_2 n_{\text{op}}(t) \end{pmatrix} \tag{S4}$$

where

$$\sigma_{nn,j}(t) = -i\kappa_{\text{in}}\, {g_j}^2\, \frac{P_n(t)}{\hbar\Omega} \left|\chi_c(\Delta_n(t))\right|^2 \left(\chi_c\left(\omega_j^{(0)} + \Delta_{\text{n}}(t)\right) - \chi_c\left(\omega_j^{(0)} - \Delta_n(t)\right)\right)$$

$$\sigma_{12}(t) = -i\kappa_{\text{in}}\, g_1 g_2 \frac{\sqrt{P_1(t)P_2(t)}}{\hbar\Omega} {\chi_c}^*(\Delta_1(t))\chi_c(\Delta_2(t)) \left(\chi_c\left(\omega_1^{(0)} + \Delta_1(t)\right) - \chi_c\left(\omega_1^{(0)} - \Delta_2(t)\right)\right)$$

$$\sigma_{21}(t) = -i\kappa_{\text{in}}\, g_1 g_2 \frac{\sqrt{P_1(t)P_2(t)}}{\hbar\Omega} {\chi_c}^*(\Delta_2(t))\chi_c(\Delta_1(t)) \left(\chi_c\left(\omega_2^{(0)} + \Delta_2(t)\right) - \chi_c\left(\omega_2^{(0)} - \Delta_1(t)\right)\right)$$

$$n_{\text{op}}(t) = \kappa_{\text{in}} \left(\frac{P_1(t)}{\hbar\Omega} |\chi_c(\Delta_1(t))|^2 + \frac{P_2(t)}{\hbar\Omega} |\chi_c(\Delta_2(t))|^2\right)$$

$$\Delta_{12}(t) = \Delta_1(t) - \Delta_2(t)$$

$$\theta_{12}(t) = \theta_1(\mathrm{t}) - \theta_2(\mathrm{t}) \tag{S5}$$

and $\chi_c(\omega) = (\kappa/2 - i\omega)^{-1}$ is the optical cavity susceptibility.

### *1.2 The rotating frame $\mathcal{R}$*

We note that $H_{\mathrm{lab}}$ is explicitly time dependent even if all of its parameters are constant, owing to the intracavity beatnote that appears in its off-diagonal terms, $e^{\pm i(\Delta_{12}t)}$. For simplicity, we remove this time dependence (but not the time dependence resulting from variations of the control parameters $P_n, \Delta_n, \theta_n$) by applying a unitary transformation $S_{\mathrm{R}}$ :

$$S_{\mathrm{R}}(t) = \begin{pmatrix} e^{i\left(-\Delta_{12}+\omega_1^{(0)}+\omega_2^{(0)}\right)t/2} & 0 \\ 0 & e^{i(\Delta_{12}+\omega_1^{(0)}+\omega_2^{(0)})t/2} \end{pmatrix} \tag{S6}$$

Equivalently,

$$S_{\mathrm{R}}(t) = \begin{pmatrix} e^{i\left(\omega_1^{(0)}+\frac{\eta}{2}\right)t} & 0 \\ 0 & e^{i\left(\omega_2^{(0)}-\frac{\eta}{2}\right)t} \end{pmatrix} \tag{S7}$$

where $\eta = \omega_2^{(0)} - \omega_1^{(0)} - \Delta_{12}$. In this frame $\mathcal{R}$, the equations of motion are:

$$i\dot{\vec{c}}(t) = H(t)\vec{c}(t) \tag{S8}$$

where

$$\vec{c}(t) = \begin{pmatrix} c_1(t) \\ c_2(t) \end{pmatrix} = S_{\mathrm{R}}(t)\vec{c}_{\mathrm{lab}}(t) \tag{S9}$$

and

$$H(t) = S_{\mathrm{R}}(t) H_{\mathrm{lab}}(t) S_{\mathrm{R}}(t)^{-1} + i\dot{S}_{\mathrm{R}}(t) S_{\mathrm{R}}(t)^{-1} \tag{S10}$$

Under this simplification, the time-dependence of $H(t)$ is solely contained in the time dependence of the control parameters $P_n(t), \Delta_n(t), \theta_n(t)$:

$$H(t) = \begin{pmatrix} -\frac{\eta}{2} - \frac{i\gamma_1^{(0)}}{2} + \sum_{n=1}^{2} \sigma_{nn,1}(t) + A_1 n_{\mathrm{op}}(t) & \sigma_{12}(t) e^{i\theta_{12}(t)} \\ \sigma_{21}(t) e^{-i\theta_{12}(t)} & \frac{\eta}{2} - \frac{i\gamma_2^{(0)}}{2} + \sum_{n=1}^{2} \sigma_{nn,2}(t) + A_2 n_{\mathrm{op}}(t) \end{pmatrix} \tag{S11}$$

Upon diagonalization of $H$, we obtain its eigenvalues $\vec{\Lambda} = (\lambda_+, \lambda_-)$ in the frame $\mathcal{R}$. In the lab frame, the mechanical susceptibility will contain components [3] near both $\omega_1^{(0)}$ and $\omega_2^{(0)}$. Thus, it is convenient to define the quantities

$$\begin{aligned} \vec{\Lambda}_1 &= (\Lambda_{1+}, \Lambda_{1-}) = \left(\omega_1^{(0)} + \frac{\eta}{2} + \lambda_+, \omega_1^{(0)} + \frac{\eta}{2} + \lambda_-\right) = \omega_1^{(0)} + \frac{\eta}{2} + \vec{\Lambda} \\ \vec{\Lambda}_2 &= (\Lambda_{2+}, \Lambda_{2-}) = \left(\omega_2^{(0)} - \frac{\eta}{2} + \lambda_+, \omega_2^{(0)} - \frac{\eta}{2} + \lambda_-\right) = \omega_2^{(0)} - \frac{\eta}{2} + \vec{\Lambda} \end{aligned} \tag{S12}$$

which will be useful when considering the mechanical response in the lab frame. In the absence of any control tones, the magnitude of (the real part of) the non-degeneracy is $|\eta|$. Note that $\vec{\Lambda}$ is independent of $\theta_{12}$.

Equation S11 may conveniently be written as:

$$H(t) = \begin{pmatrix} L(t) & M(t) e^{i\theta_{12}(t)} \\ N(t) e^{-i\theta_{12}(t)} & -L(t) \end{pmatrix} + \frac{\mathcal{T}(t)}{2} I \tag{S13}$$

While a constant $\theta_{12}$ can always be set to zero by a time-independent change of coordinates, this is not possible for time-varying $\theta_{12}$. Here $I$ is the $2 \times 2$ identity matrix, and

$$\mathcal{T}(t) = -i\left(\frac{\gamma_1^{(0)} + \gamma_2^{(0)}}{2}\right) + \sum_{n=1}^{2} \left(\sigma_{nn,1}(t) + \sigma_{nn,2}(t)\right) + (\mathrm{A}_1 + \mathrm{A}_2) n_{\mathrm{op}}(t) \tag{S14}$$

$$L(t) = -\frac{\eta}{2} - i\left(\frac{\gamma_1^{(0)} - \gamma_2^{(0)}}{4}\right) + \sum_{n=1}^{2}\left(\frac{\sigma_{nn,1}(t) - \sigma_{nn,2}(t)}{2}\right) + \left(\frac{\mathrm{A}_1 - \mathrm{A}_2}{2}\right) n_{\mathrm{op}}(t) \qquad \text{(S15)}$$

$$M(t) = \sigma_{12}(t) \qquad N(t) = \sigma_{21}(t) \qquad \text{(S16)}$$

While $\mathcal{T}$ (i.e., the trace of $H$) is irrelevant for the geometric phase of a control loop, it still impacts the measured signals. For example, $\mathrm{Im}(\mathcal{T})$ sets the modes' overall damping, which in turn sets a practical upper limit on the duration of control loops. Also, as discussed in Supplemental Note 6, $\mathrm{Im}(\mathcal{T})$ also plays an important role in producing stead-state geometric gain.

## Supplemental Note 2: System characterization

This Supplemental Note describes the measurements used to validate the model described in Supplemental Note 1, and to determine the values of the relevant parameters.

### *2.1 Undriven motion*

To extract the bare membrane parameters $\omega_i^{(0)}, \gamma_i^{(0)}$ (the frequencies and damping rates of the $i^{\text{th}}$ mode in the absence of dynamical back action), we turn the control tones off and measure the spectrum of the membrane's Brownian motion. We fit the power spectral density of these fluctuations in the vicinity of each resonance to the expected form (the square modulus of a Lorentzian plus a constant), which returns the best-fit value of each mode's complex eigenfrequency $\lambda_i^{(0)} = \omega_i^{(0)} - i\gamma_i^{(0)}/2$. A representative measurement for each mode is shown in Supplemental Figure 2a–c.

### *2.2 Characterization using a single control tone*

In presence of a single control tone, $H$ is diagonal, and its eigenvalues (i.e. $H_{11}$ and $H_{22}$) encode the change in damping rate (imaginary part) and resonance frequency (real part) for each mechanical mode. The damping rate is tuned via radiation pressure-induced dynamical back-action (DBA) [4], while the resonance frequency is tuned by both DBA and the photothermal effect defined above.

To illustrate this, we apply a single control tone with power $P_1 = 17$ μW and measure the membrane's mechanical susceptibility as a function of the control beam's detuning $\Delta_1/2\pi$. From these measurements, we extract the shift in mechanical frequency $\delta\omega_i$ and the damping $\gamma_i$. This data, along with a least-squares fit (using the eigenvalues of $H$ as fit parameters) is show in Supplemental Figure 2d–f.

These fits include an additional parameter $\Delta_{\text{off}}$ that represents the offset between the laser lock and the cavity resonance. For all the measurements described in this paper, we find $\Delta_{\text{off}}/2\pi < 50$ kHz. While this offset is essentially negligible for the analysis presented here, for consistency we replace $\Delta \rightarrow \Delta + \Delta_{\text{off}}$ in all expressions when comparing them with data.

### *2.3 Characterization using two control tones*

Here, we provide details of the susceptibility measurements shown in Fig. 1c. These measurements are similar to those just described, except we now apply two control tones, resulting in coupling between the two mechanical modes. As described in the main text, we measure two copies of the mechanical susceptibility near $\omega_1^{(0)}$ and $\omega_2^{(0)}$ while varying $P_1 = P_2 = P$ and $\delta$ on a rectangular grid with a fixed value of $\eta$. Each copy of the mechanical susceptibility is fit to the sum of two complex Lorentzians (plus a complex constant to account for feed-through from the drive) to obtain the system's eigenvalues in the rotating frame $\mathcal{R}$. In particular, the mechanical susceptibility near $\omega_1^{(0)}$ ($\omega_2^{(0)}$) is fit to extract the complex eigenfrequencies $\vec{\Lambda}_1 = (\Lambda_{1+}, \Lambda_{1-})$ ($\vec{\Lambda}_2 = (\Lambda_{2+}, \Lambda_{2-})$), as illustrated in Fig. 1c.

This fit provides $\lambda_+, \lambda_-$, from which we calculate $D = (\lambda_+ - \lambda_-)^2$ and $\mathcal{T} = \lambda_+ + \lambda_-$ at each point $(P, \delta)$ on the grid. $D$ and $\mathcal{T}$ are simple functions of $H$:

$$\begin{aligned} D &= \mathcal{T}^2 - 4\det(H) \\ \mathcal{T} &= \mathrm{tr}(H) \end{aligned} \tag{S17}$$

where tr is the trace, det is the determinant, and $D$ is the discriminant of the matrix's characteristic polynomial. Using the above relations, we fit the measured $D$ and $\mathcal{T}$ to obtain the best-fit values of $\kappa, g_1, g_2, A_1, A_2, \Delta_{\mathrm{off}}$, which define the optomechanical Hamiltonian of our system. These values are given in Supplemental Table 1.

## Supplemental Note 3: Measuring the propagator matrix for a control path

This Supplemental Note gives a detailed description of the protocol used to measure the phase accumulated by the membrane's oscillations when its control parameters are varied along a path.

### *3.1 Protocol for measuring $\beta(T)$*

In this section, we provide additional information on measuring $\beta(T)$ for a given control path. The process of measuring $\beta(T)$ is summarized in the following steps:

1. The optomechanical control parameters $(P_1, P_2, \delta, \eta, \theta_{12})$ are tuned to their values at the beginning of the loop, setting the Hamiltonian of the mechanical system to its initial value $H(0)$.
2. A drive is applied at frequency $\omega_{\mathrm{d}}^{(a)}$ to ring the membrane up to its initial state vector $\vec{c}^{(a)}(0)$.
3. The drive is turned off. This defines $t = 0$.
4. While the membrane rings down (i.e. evolves under the time independent Hamiltonian $H(0)$), we record the heterodyne signal demodulated by two oscillators at frequencies $\omega_1^{\mathrm{mod}} = \omega_{\mathrm{d}}^{(a)}$ and $\omega_2^{\mathrm{mod}}$. This measurement corresponds to the left column of Fig. 2c,d.
5. Steps 2 – 3 are repeated. After the drive is turned off, the control parameters are tuned along some control path for a duration $T$ (e.g., by setting $\theta_{12}(t) = 2\pi t/T$) such that the system evolves under a time dependent Hamiltonian $H(t)$. For $t > T$ the membrane is allowed to ring down (i.e. evolve under constant $H(T)$), and we record the heterodyne signal demodulated by the same two oscillators at $\omega_1^{\mathrm{mod}}$ and $\omega_2^{\mathrm{mod}}$. This measurement corresponds to the right column of Fig. 2c,d.
6. We repeat the measurements described above, alternating between free ringdown (step 4) and the control path (step 5). Typically 50 – 500 measurements of each type are averaged to increase the signal to noise ratio (SNR).
7. We fit the "free ringdown" data to extract the initial state vector $\vec{c}^{(a)}(0)$ immediately after the drive is turned off, and the "control path" data to extract the state vector $\vec{c}^{(a)}(T)$ after a control path of duration $T$ is performed (see Supplemental Note 3.2 for details).
8. In order to determine the four complex components of the propagator matrix $U(T)$ (defined by $\vec{c}(T) = U(T)\vec{c}(0)$), we measure the evolution of two linearly independent initial state vectors. Thus, we repeat steps 2 – 7 above, choosing a different value for the drive frequency ($\omega_{\mathrm{d}}^{(b)}$ instead of $\omega_{\mathrm{d}}^{(a)}$), which rings the membrane up to an initial state vector $\vec{c}^{(b)}(0)$ that is linearly independent of $\vec{c}^{(a)}(0)$. The vectors $(\vec{c}^{(a)}(0), \vec{c}^{(b)}(0), \vec{c}^{(a)}(T), \vec{c}^{(b)}(T))$ yield sufficient information to determine the propagator in the "forwards" direction $U_{\circlearrowleft}(T)$ (see Supplemental Note 3.2 for details).
9. We repeat the entire series of measurements described in steps 2 – 8, varying the control path duration $T$ to determine $U_{\circlearrowleft}(T)$ as a function of $T$.

10. We repeat the series of measurements described in steps 1 – 9, except that we apply the "time reversed" version of the control path from the above measurements. For example, if the "forward" circuit was defined by $\theta_{12}(t) = 2\pi t/T$, the "time reversed" circuit would be defined by $\theta_{12}(t) = -2\pi t/T$ (see Fig. 2c, magenta line). This measurement gives the "time reversed" propagator $U_{\circlearrowright}(T)$.
11. From these data, we calculate $\beta(T) = -i\log(U_{++,\circlearrowleft}/U_{++,\circlearrowright})/2$ which (up to a choice of the branch of the logarithm, see discussion in Supplemental Note 4) tends to the complex geometric phase $\phi_{\mathrm{B},\circlearrowright}$ at large $T$. Here the least-damped mode is indexed by the subscript $+$. This sequence constitutes the measurement of $\beta(T)$ for a given control path.

Prior to performing this measurement sequence, we measure the Brownian motion spectrum with the control tones off to obtain the bare mechanical frequencies ($\omega_1^{(0)}$ and $\omega_2^{(0)}$, see Supplemental Note 2). We then perform a measurement of the membrane's susceptibility with the control tones on to extract the frequencies $\Lambda_{ij}$. This static spectroscopy is also performed intermittently during a measurement sequence to track any drift in the setup's parameters that may occur during data acquisition. A typical measurement of $\beta(T)$ for a given control loop (see for example Fig. 2g) lasts ~ 4 hours, during which time the static spectroscopy measurement is performed every 15 minutes. The drive and demodulation frequencies (discussed in the following paragraphs) are updated based on these measurements.

The values of $\Lambda_{ij}$ gleaned from these measurements are used to set the drive frequency $\omega_{\mathrm{d}}$ and the demodulation frequencies $\omega_1^{\mathrm{mod}}$ and $\omega_2^{\mathrm{mod}}$. Specifically, we choose $\omega_{\mathrm{d}}$ to be the real part of one of the four $\Lambda_{ij}$. Any choice of the four values of $\mathrm{Re}(\Lambda_{ij})$ results in a distinct initialization of the mechanical state vector. However, to reconstruct the propagator matrix $U(T)$ for a control path of duration $T$, we measure initial and the final state vectors for two linearly independent initializations of the state vector; these are the $\vec{c}^{(a)}(0)$ and $\vec{c}^{(b)}(0)$ introduced above in Steps 2 & 8. To ensure that $\vec{c}^{(a)}(0)$ and $\vec{c}^{(b)}(0)$ are linearly independent to a degree sufficient to accurately determine the propagator, we initialize $\vec{c}^{(a)}(0)$ with $\omega_{\mathrm{d}}^{(a)} = \mathrm{Re}(\Lambda_{1j}) \sim \omega_1^{(0)}$ and $\vec{c}^{(b)}(0)$ with $\omega_{\mathrm{d}}^{(b)} = \mathrm{Re}(\Lambda_{2j}) \sim \omega_2^{(0)}$.

The frequencies of the demodulation oscillators are also chosen from among the four $\mathrm{Re}(\Lambda_{ij})$ with one being equal to $\omega_{\mathrm{d}}$. Specifically, we choose $\omega_1^{\mathrm{mod}} = \mathrm{Re}(\Lambda_{1j}) \sim \omega_1^{(0)}$ and $\omega_2^{\mathrm{mod}} = \mathrm{Re}(\Lambda_{2j}) \sim \omega_2^{(0)}$.

When averaging multiple records of the heterodyne signal $V(t)$ (with the $i$th record denoted as $V_i(t)$), it is important to account for the phase of $V_i(t)$. For each $V_i(t)$, we reference the phase at the end of the drive to zero, i.e. we set $\arg[V_i(0)] = 0$. This ensures that, upon averaging, coherent motion of the membrane induced by the drive is in phase for all $i$, while incoherent noise will tend to average to zero. It is also important to choose the same reference time (i.e., the end of the drive) for both the "free ringdown" measurement and the "control path" measurement, as this ensures that the state vectors $\vec{c}^{(a)}(0)$ and $\vec{c}^{(a)}(T)$ have a common phase reference. The averaging is illustrated in Supplemental Figure 3b, where individual measurements (faint lines) are plotted along with the averaged result (thick green line).

### *3.2 Extracting the state vector from ringdown measurements*

When the control parameters are held fixed and the drive is off, the free ringdown of the mechanical modes may be written in the lab frame as a solution $\vec{c}_{\mathrm{lab}}(t)$ to Eq. S3. For simplicity, we first solve for $\vec{c}(t)$ in the rotating frame $\mathcal{R}$ and then apply a unitary transformation ($S_{\mathrm{R}}^{-1}$) to convert motional amplitudes in the frame $\mathcal{R}$ to the lab frame (see Supplemental Note 1).

To solve $\vec{c}(t)$ in the frame $\mathcal{R}$, we rewrite Eq. S3 as:

$$i\dot{\vec{c}}_{\mathrm{D}}(t) = H_{\mathrm{D}}\vec{c}_{\mathrm{D}}(t) \tag{S18}$$

where we introduce the diagonal frame state vector $\vec{c}_{\mathrm{D}}(t) = \left(\bar{c}_{+}(t), \bar{c}_{-}(t)\right)^{T} = S_{\mathrm{D}}^{-1}\vec{c}(t)$ and Hamiltonian

$$H_{\mathrm{D}} = \begin{pmatrix} \lambda_{+} & 0 \\ 0 & \lambda_{-} \end{pmatrix} = S_{\mathrm{D}}{}^{-1} H S_{\mathrm{D}} \tag{S19}$$

where $\lambda_{+}$ and $\lambda_{-}$ are the eigenvalues of the system in frame $\mathcal{R}$ and the columns of $S_{\mathrm{D}}$ (rows of $S_{\mathrm{D}}{}^{-1}$) are composed of the normalized right (left) eigenvectors of $H$. We emphasize that $H, H_{\mathrm{D}}, S_{\mathrm{D}}, S_{\mathrm{D}}^{-1}$ are time independent because the control parameters are held fixed before the start of the loop (for state initialization), and at the end of the loop (for post-control-loop ringdown).

The solution of Eq. S18 is:

$$\vec{c}_{\mathrm{D}}(t) = \begin{pmatrix} \bar{c}_{+}e^{-i\lambda_{+}t} \\ \bar{c}_{-}e^{-i\lambda_{-}t} \end{pmatrix} \tag{S20}$$

where $\bar{c}_{i}$ is the complex motional amplitude of normal mode $i \in \{+,-\}$ at the instant when the radiation pressure drive is turned off. Writing

$$S_{\mathrm{D}} = \begin{pmatrix} u_{11} & u_{12} \\ u_{21} & u_{22} \end{pmatrix} \tag{S21}$$

gives

$$\vec{c}(t) = S_{\mathrm{D}}\vec{c}_{\mathrm{D}}(t) = \begin{pmatrix} u_{11}\bar{c}_{+}e^{-i\lambda_{+}t} + u_{12}\bar{c}_{-}e^{-i\lambda_{-}t} \\ u_{21}\bar{c}_{+}e^{-i\lambda_{+}t} + u_{22}\bar{c}_{-}e^{-i\lambda_{-}t} \end{pmatrix} \tag{S22}$$

and

$$\vec{c}_{\text{lab}}(t) = S_{\text{R}}^{-1}(t)\vec{c}(t) = S_{\text{R}}^{-1}(t)S_{\text{D}}\vec{c}_{\text{D}}(t) = \begin{pmatrix} u_{11}\bar{c}_+ e^{-i\left(\lambda_+ + \omega_1^{(0)} + \frac{\eta}{2}\right)t} + u_{12}\bar{c}_- e^{-i\left(\lambda_- + \omega_1^{(0)} + \frac{\eta}{2}\right)t} \\ u_{21}\bar{c}_+ e^{-i\left(\lambda_+ + \omega_2^{(0)} - \frac{\eta}{2}\right)t} + u_{22}\bar{c}_- e^{-i\left(\lambda_- + \omega_2^{(0)} - \frac{\eta}{2}\right)t} \end{pmatrix} \quad \text{(S23)}$$

Or, more succinctly via Eq. S12:

$$\vec{c}_{\text{lab}}(t) = \begin{pmatrix} u_{11}\bar{c}_+ e^{-i\Lambda_{1+}t} + u_{12}\bar{c}_- e^{-i\Lambda_{1-}t} \\ u_{21}\bar{c}_+ e^{-i\Lambda_{2+}t} + u_{22}\bar{c}_- e^{-i\Lambda_{2-}t} \end{pmatrix} \quad \text{(S24)}$$

This motion is transduced onto the optical field, converted into an electronic signal, and measured by the LIA as $\rho\chi_{\text{c}}(\omega)\vec{g}\cdot\vec{c}_{\text{lab}}(t)$, where $\rho$ is the transduction gain, $\chi_{\text{c}}(\omega)$ is the cavity susceptibility and $\vec{g} = (g_1, g_2)$ is the vector of optomechanical coupling strengths and $\cdot$ is the usual dot product. This signal is then demodulated by the LIA at frequencies $\omega_1^{\text{mod}}$ and $\omega_2^{\text{mod}}$ and passed through a low-pass filter with bandwidth $BW \ll \left|\omega_1^{(0)} - \omega_2^{(0)}\right|$. The complex demodulated signals at $\omega_1^{\text{mod}}$ and $\omega_2^{\text{mod}}$ are:

$$\begin{aligned} V_1(t) &= \rho\chi_{\text{c}}\left(\omega_1^{\text{mod}}\right)g_1\left(W\left(\Lambda_{1+}, \omega_1^{\text{mod}}\right) u_{11}\bar{c}_+ e^{-i\Lambda_{1+}t} + W\left(\Lambda_{1-}, \omega_1^{\text{mod}}\right)u_{12}\bar{c}_- e^{-i\Lambda_{1-}t}\right)e^{i\omega_1^{\text{mod}}t} \\ V_2(t) &= \rho\chi_{\text{c}}\left(\omega_2^{\text{mod}}\right)g_2\left(W\left(\Lambda_{2+}, \omega_2^{\text{mod}}\right) u_{21}\bar{c}_+ e^{-i\Lambda_{2+}t} + W\left(\Lambda_{2-}, \omega_2^{\text{mod}}\right)u_{22}\bar{c}_- e^{-i\Lambda_{2-}t}\right)e^{i\omega_2^{\text{mod}}t} \end{aligned} \quad \text{(S25)}$$

respectively, where $W(\lambda, \omega) = (1 + i(\text{Re}(\lambda) - \omega)\tau_{\text{BW}})^{-n}$ is the low-pass filter function for a signal at frequency $\text{Re}(\lambda)$ that is demodulated by an oscillator at $\omega$ with a low-pass filter whose bandwidth, time-constant, and order are denoted as $BW, \tau_{\text{BW}}$ and $n$ respectively. For the measurements in this manuscript, $n = 1$, $BW = 250$ Hz, and $\tau_{\text{BW}} = (2\pi BW)^{-1} = 0.637$ ms.

In addition, the LIA has a settling time $\tau_{\text{settle}}$ (corresponding to the 99% settling time of the LIA) that depends on $n$ and $\tau_{\text{BW}}$. For these measurements, $\tau_{\text{settle}} = 2.934$ ms. To accommodate this, we exclude from any fits the data in an interval $\tau_{\text{settle}}$ after the start of a ringdown.

Both $V_1(t)$ and $V_2(t)$ contain equivalent information about the complex motional amplitudes $\bar{c}_+$ and $\bar{c}_-$, up to known multiplicative factors ($\chi_{\text{c}}(\omega), W(\lambda, \omega), g_i, u_{ij}$) and an overall scaling factor $\rho$. As a reminder, the motional eigenstates at $t = 0$ are extracted from measurements of "simple" ringdowns, i.e. ringdowns without a control path. The signals $V_1(t)$ and $V_2(t)$ recorded in such a measurement (Eq. S25) are fit to:

$$\begin{aligned} V_1(t) &= b_1 + A_{11}(t)\, W\left(\Lambda_{1+}, \omega_1^{\text{mod}}\right)e^{-i\left(\Lambda_{1+} - \omega_1^{\text{mod}}\right)t} + A_{12}(t)\, W\left(\Lambda_{1-}, \omega_1^{\text{mod}}\right)e^{-i\left(\Lambda_{1-} - \omega_1^{\text{mod}}\right)t} \\ V_2(t) &= b_2 + A_{21}(t)\, W\left(\Lambda_{2+}, \omega_2^{\text{mod}}\right)e^{-i\left(\Lambda_{2+} - \omega_2^{\text{mod}}\right)t} + A_{22}(t)\, W\left(\Lambda_{2-}, \omega_2^{\text{mod}}\right)e^{-i\left(\Lambda_{2-} - \omega_2^{\text{mod}}\right)t} \end{aligned} \quad \text{(S26)}$$

respectively. Here, $\lambda_\pm$ and $b_i, A_{ij}$, with $i, j \in 1,2$ are the 8 complex-valued fit parameters, which denote the system's eigenvalues, the heterodyne signal's background, and the amplitudes of the decaying exponentials, respectively ($\lambda_\pm$ appear in Eq. S26 by way of Eq. S12). By comparing Eqs. S25 and S26, the complex amplitudes at $t = 0$ can be written as

$$
\begin{aligned}
\bar{c}_+(0) &= \frac{A_{11}(0)}{\rho\chi_c\left(\omega_1^{(0)}\right) g_1 u_{11}} = \frac{A_{21}(0)}{\rho\chi_c\left(\omega_2^{(0)}\right) g_2 u_{21}} \\
\bar{c}_-(0) &= \frac{A_{12}(0)}{\rho\chi_c\left(\omega_1^{(0)}\right) g_1 u_{12}} = \frac{A_{22}(0)}{\rho\chi_c\left(\omega_2^{(0)}\right) g_2 u_{22}}
\end{aligned} \tag{S27}
$$

which highlights the fact $V_1(t)$ and $V_2(t)$ contain redundant information about motional amplitudes. In practice, we simply use the fit parameters obtained from the fit to $V_1(t)$ (i.e. $A_{11}(t)$ and $A_{12}(t)$) to calculate $\bar{c}_+(0)$ and $\bar{c}_-(0)$ (this is justified in the next paragraph).

Expressions similar to Eqs. S27 may also be used to extract the amplitudes of the normal modes' motion at $t = T$ i.e., $\bar{c}_+(T)$ and $\bar{c}_-(T)$, from the data corresponding to ringdown after a control path. However, for large $T$, the motion may decay significantly before the completion of the control path, resulting in low SNR and potentially compromising the quality of the fit. In practice, we observe that the signal near $\omega_1^{(0)}$, and correspondingly $V_1(t)$, tends to have a larger SNR than $V_2(t)$ near $\omega_2^{(0)}$ after long loops. Thus, to obtain complex motional amplitudes at the end of the loop, we fit only $V_1(t)$.

Additionally, we use only 3 complex fit parameters $(b_i, A_{11}(T), A_{12}(T))$ in this fit, fixing the eigenvalues $\lambda_1$ and $\lambda_2$ to values obtained by fitting the corresponding initialization ringdown data. We justify the choice of fixing the eigenvalues during the fit by noting that (a) for all control loops in this manuscript, the eigenvalues at the beginning of the circuit are equal to the eigenvalues at the end of the circuit, and (b) the individual measurements of complex-averaged datasets were interleaved, which reduces the impact of systematic errors caused by temporal drifts.

From this fit, we obtain the complex motional amplitudes at $t = T$ which can be written as

$$
\bar{c}_+(T) = \frac{A_{11}(T)}{\rho\chi_c\left(\omega_1^{(0)}\right) g_1 u_{11}} \tag{S28}
$$

$$
\bar{c}_-(T) = \frac{A_{12}(T)}{\rho\chi_c\left(\omega_1^{(0)}\right) g_1 u_{12}} \tag{S29}
$$

### *3.3 Measuring the propagator matrix* $\boldsymbol{U(T)}$

As described above, in order to reconstruct the full propagator $U(T)$ for a control path of duration $T$, we measure initial and the final state vectors for two linearly independent initializations of the state vector $\vec{c}^{(a)}(0)$ and $\vec{c}^{(b)}(0)$. Thus, for every value of $T$, we measure $\left\{\bar{c}_+^{(a)}(0), \bar{c}_-^{(a)}(0), \bar{c}_+^{(a)}(T), \bar{c}_-^{(a)}(T)\right\}$ and $\left\{\bar{c}_+^{(b)}(0), \bar{c}_-^{(b)}(0), \bar{c}_+^{(b)}(T), \bar{c}_-^{(b)}(T)\right\}$. The matrix equation that connects the initial and final complex amplitudes is given by:

$$\begin{pmatrix} \bar{c}_+(T) \\ \bar{c}_-(T) \end{pmatrix} = U(T) \begin{pmatrix} \bar{c}_+(0) \\ \bar{c}_-(0) \end{pmatrix} \tag{S30}$$

where $U(T)$ is the complex valued propagator matrix. Explicitly,

$$U(T) = \begin{pmatrix} U_{++}(T) & U_{+-}(T) \\ U_{-+}(T) & U_{--}(T) \end{pmatrix} \tag{S31}$$

The two initializations $\vec{c}^{(a)}(0)$ and $\vec{c}^{(b)}(0)$ result in four linearly independent equations relating the initial and final complex amplitudes:

$$\begin{aligned}
\bar{c}_+^{(a)}(T) &= U_{++}(T)\bar{c}_+^{(a)}(0) + U_{+-}(T)\bar{c}_-^{(a)}(0) \\
\bar{c}_+^{(b)}(T) &= U_{++}(T)\bar{c}_+^{(b)}(0) + U_{+-}(T)\bar{c}_-^{(b)}(0) \\
\bar{c}_-^{(a)}(T) &= U_{-+}(T)\bar{c}_+^{(a)}(0) + U_{--}(T)\bar{c}_-^{(a)}(0) \\
\bar{c}_-^{(b)}(T) &= U_{-+}(T)\bar{c}_+^{(b)}(0) + U_{--}(T)\bar{c}_-^{(b)}(0)
\end{aligned} \tag{S32}$$

These four equations are sufficient to solve for $U(T)$, whose elements are given by:

$$U_{++}(T) = \frac{\bar{c}_+^{(a)}(T)\bar{c}_{-,n}(0) - \bar{c}_+^{(b)}(T)\bar{c}_-^{(a)}(0)}{\bar{c}_+^{(a)}(0)\bar{c}_-^{(b)}(0) - \bar{c}_+^{(b)}(0)\bar{c}_-^{(a)}(0)}$$

$$U_{+-}(T) = \frac{\bar{c}_+^{(a)}(T)\bar{c}_+^{(b)}(0) - \bar{c}_+^{(b)}(T)\bar{c}_+^{(a)}(0)}{\bar{c}_-^{(a)}(0)\bar{c}_+^{(b)}(0) - \bar{c}_-^{(b)}(0)\bar{c}_+^{(a)}(0)}$$

$$U_{-+}(T) = \frac{\bar{c}_{-,m}(T)\bar{c}_-^{(b)}(0) - \bar{c}_-^{(b)}(T)\bar{c}_-^{(a)}(0)}{\bar{c}_+^{(a)}(0)\bar{c}_-^{(b)}(0) - \bar{c}_+^{(b)}(0)\bar{c}_-^{(a)}(0)}$$

$$U_{--}(T) = \frac{\bar{c}_{-}^{(a)}(T)\bar{c}_{+}^{(b)}(0) - \bar{c}_{-}^{(b)}(T)\bar{c}_{+}^{(a)}(0)}{\bar{c}_{-}^{(a)}(0)\bar{c}_{+}^{(b)}(0) - \bar{c}_{-}^{(b)}(0)\bar{c}_{+}^{(a)}(0)} \tag{S33}$$

Using the relation between the motional amplitude and the amplitudes of the heterodyne signal (Eqs. 27 – 29), these expressions can be simplified to:

$$U_{++}(T) = \left(\frac{A_{11}^{(a)}(T)A_{12}^{(b)}(0) - A_{11}^{(b)}(T)A_{12}^{(a)}(0)}{A_{11}^{(a)}(0)A_{12}^{(b)}(0) - A_{11}^{(b)}(0)A_{12}^{(a)}(0)}\right) . \frac{u_{11}(0)}{u_{11}(T)}$$

$$U_{+-}(T) = \left(\frac{A_{11}^{(a)}(T)A_{11}^{(b)}(0) - A_{11}^{(b)}(T)A_{11}^{(a)}(0)}{A_{12}^{(a)}(0)A_{11}^{(b)}(0) - A_{12}^{(b)}(0)A_{11}^{(a)}(0)}\right) . \frac{u_{12}(0)}{u_{11}(T)}$$

$$U_{-+}(T) = \left(\frac{A_{12}^{(a)}(T)A_{12}^{(b)}(0) - A_{12}^{(b)}(T)A_{12}^{(a)}(0)}{A_{11}^{(a)}(0)A_{12}^{(b)}(0) - A_{11}^{(b)}(0)A_{12}^{(a)}(0)}\right) . \frac{u_{11}(0)}{u_{12}(T)}$$

$$U_{--}(T) = \left(\frac{A_{12}^{(a)}(T)A_{11}^{(b)}(0) - A_{12}^{(b)}(T)A_{11}^{(a)}(0)}{A_{12}^{(a)}(0)A_{11}^{(b)}(0) - A_{12}^{(b)}(0)A_{11}^{(a)}(0)}\right) . \frac{u_{12}(0)}{u_{12}(T)} \tag{S34}$$

where the $A_{ij}^{(k)}$ are the amplitudes of motion at $t = 0$ or $t = T$ returned by fitting the ringdown of the heterodyne signal (see Supplemental Note 3.2) and the $u_{ij}$ are given in Eq. S21.

Notably, the propagator matrix $U(T)$ depends only on the $A_{ij}^{(k)}$ and the choice of gauge for the eigenvectors that compose $S_{\mathrm{D}}^{-1}$. Since the adiabatic theorem is only applicable to the mode with least loss [5], only the diagonal component corresponding to the least-damped mode (denoted by subscript $++$) contains information about the geometric phase. For closed loops, we choose $u_{ij}(0) = u_{ij}(T)$ (since $t = 0$ and $t = T$ correspond to the same point in parameter space), with the result that $U_{++}(T)$ and $U_{--}(T)$ (Eq. S34) are independent of the $u_{ij}$.

While $U_{++}(T)$ is the main focus of this paper, for completeness Supplemental Figures 3c-f show an example measurement of all four propagator matrix elements for control loops with $P_1 = P_2 = 15$ μW , $\delta/2\pi = -1$ MHz, $\eta/2\pi = -50$ Hz, and $\theta_{12}(t) = \pm 2\pi t/T$ (as also used in Fig. 2). The off-diagonal elements $U_{+-}(T)$ and $U_{-+}(T)$ use the $u_{ij}$ calculated from the optomechanical model, and in a gauge where $u_{ij} \in \mathbb{R}$.

The absence of an adiabatic limit for the more-damped mode is evident in the fact that $|U_{--}(T)| \ll |U_{+-}(T)|$ for large $T$. This is visible in the data and theory shown Supplemental Figure 3d, and in the theory in Supplemental Figure 3c.

**Supplemental Note 4. The propagator matrix and the geometric phase**

This Supplemental Note summarizes some important features of the phase that is accumulated by a state when it is adiabatically transported. In particular, we review the representation of this phase as a power series in $1/T$; the impact of reversing the direction in which a given control path is traversed; and important features related to the definition of (real) phases modulo 2π.

***4.1 The asymptotic form of the accumulated phase***

We start with a time dependent (potentially with non-zero trace) Hamiltonian $H(t)$ which is diagonalized by a time dependent change of basis $S_{\mathrm{D}}(t)$, i.e. $\mathrm{H}_{\mathrm{D}}(t) = S_{\mathrm{D}}^{-1}(t)H(t)S_{\mathrm{D}}(t)$ where

$$H_{\mathrm{D}}(t) = \begin{pmatrix} \lambda_+(t) & 0 \\ 0 & \lambda_-(t) \end{pmatrix}. \tag{S35}$$

We require the column vectors of $S_{\mathrm{D}}(t)$ to be normalized (see discussion in Supplemental Note 5) and that $S_{\mathrm{D}}(t)$ is $T$-periodic. In this diagonal basis, Eq. S8 becomes

$$i\epsilon\, \partial_s \vec{c}(s) = \big(\mathrm{H}_{\mathrm{D}}(s) - \epsilon\mathcal{A}(s)\big)\vec{c}(s) \tag{S36}$$

where $s = t/T$, and we have defined $\epsilon = 1/T$ and $\mathcal{A}_{ij} = \vec{\xi}_i(s) \cdot i\, \partial_s \vec{\psi}_j(s)$ for $i,j = +,-$ where $\vec{\xi}_i$ and $\vec{\psi}_j$ are the left- and right-eigenvectors of $H$, respectively.

Now, consider a closed path through the space of Hamiltonian parameters where $\mathrm{Im}[\lambda_+(s)] < \mathrm{Im}[\lambda_-(s)]$ for all $s \in [0,1]$, i.e., $\vec{\psi}_+(s)$ decays more slowly than $\vec{\psi}_-(s)$. The adiabatic theorem for non-Hermitian Hamiltonians then guarantees that if a state is initialized in $\vec{\psi}_+(0)$ it will remain in $\vec{\psi}_+(s)$ for all $s \in [0,1]$ as $\epsilon \to 0$ [5]. Therefore, as $\epsilon \to 0$, we may use time-independent perturbation theory to find the eigenvalue $\lambda'_+(s)$ of $\mathrm{H}_{\mathrm{D}}(s) - \epsilon\mathcal{A}(s)$ at every $s$. Note that $\lambda'_+(s) \to \lambda_+(s)$ as $\epsilon \to 0$. In this case, the phase accumulated by a state vector $\vec{c}(T) = e^{-i\phi_{\circlearrowleft}}\vec{c}(0)$ (with $\vec{c}(0) = \vec{\psi}_+$) can be expressed as a power series in $\epsilon$:

$$\begin{aligned}\phi_{\circlearrowleft} = \frac{1}{\epsilon}\int_0^1 ds\, \lambda'_+(s) \\ = \int_0^1 ds \Bigg[ (1/\epsilon)\lambda_+(s) - i\vec{\xi}_+(s) \cdot \partial_s \vec{\psi}_+(s) \\ - \epsilon \frac{\big(\vec{\xi}_-(s) \cdot \partial_s \vec{\psi}_+(s)\big)\big(\vec{\xi}_+(s) \cdot \partial_s \vec{\psi}_-(s)\big)}{\lambda_+(s) - \lambda_-(s)} + \mathcal{O}(\epsilon^2) \Bigg].\end{aligned} \tag{S37}$$

The first term in this expression is the dynamical phase $\phi_{\mathrm{D}}$; the second term is the geometric phase $\phi_{\mathrm{B}}$; and the subsequent terms approach 0 in the adiabatic limit $\epsilon \to 0$. In general, the $N^{\mathrm{th}}$ term is proportional to $\epsilon^{N-2}$ and to a sum of products involving $N-1$ components of $\mathcal{A}_{ij}$ (i.e., containing $N-1$ derivatives).

To consider the phase $\phi_{\circlearrowright}$ accumulated by a time reversed loop, replace $\vec{\psi}_i(s), \vec{\xi}_i(s), \lambda_i(s)$ respectively by $\vec{\psi}_i(1-s), \vec{\xi}_i(1-s), \lambda_i(1-s)$ in the preceding formulae. In each term, each derivative acquires a factor $(-1)$ but otherwise is unchanged. Thus, the *odd* order (in $\epsilon$) terms of $\phi_{\circlearrowright}$ will be equal to the corresponding terms of $\phi_{\circlearrowleft}$, while the *even* order terms of $\phi_{\circlearrowright}$ will be equal to the corresponding terms of $\phi_{\circlearrowleft}$ multiplied by $-1$.

As a result, in the large $T$ limit we may write the phases accumulated along each direction of a control path as

$$\phi_{\circlearrowleft} = 2\pi n_{\circlearrowleft} + q_{\mathrm{D}} T - \phi_{\mathrm{B}} + \frac{q_1}{T} + \frac{q_2}{T^2} + \mathcal{O}(T^{-3}) \tag{S38}$$

$$\phi_{\circlearrowright} = 2\pi n_{\circlearrowright} + q_{\mathrm{D}} T + \phi_{\mathrm{B}} + \frac{q_1}{T} - \frac{q_2}{T^2} + \mathcal{O}(T^{-3}) \tag{S39}$$

In these expressions, $n_{\circlearrowleft}$ and $n_{\circlearrowright}$ are arbitrary integers, reflecting the fact that phases are defined modulo $2\pi$. Equivalently, the choice of $n_{\circlearrowleft,\circlearrowright}$ corresponds to the choice of a branch of the logarithm in $\widetilde{\phi}(T) = -i\log(U_{++}(T))$. We describe our convention for this choice in the following.

The quantity $\beta(T) = (\phi_{\circlearrowright} - \phi_{\circlearrowleft})/2$ contains only terms of even order in $\epsilon$ (and hence in $T$). The first three terms in the large-$T$ (small- $\epsilon$) expansion of $\beta(T)$ are of the form

$$\beta(T) = \pi(n_{\circlearrowright} - n_{\circlearrowleft}) + \phi_{\mathrm{B}} - \frac{q_2}{T^2} + \mathcal{O}(T^{-4}) \tag{S40}$$

Ignoring the first term, this is the fit function used to determine $\phi_{\mathrm{B}}$ from the data (as described in Supplemental Note 3) using the complex coefficients $q_2$ and $\phi_{\mathrm{B}}$ as fit parameters. This shows that without further processing, $\phi_{\mathrm{B}}$ can only be estimated modulo $\pi$ from $\beta(T)$.

However, $\phi_{\mathrm{B}}$ may be determined modulo $2\pi$ by inspecting the quantity $\alpha(T) = \phi_{\circlearrowright} + \phi_{\circlearrowleft}$, which contains only terms of odd order in $T$:

$$\mathrm{Re}(\alpha(T)) = 2\pi(n_{\circlearrowright} + n_{\circlearrowleft}) + 2\mathrm{Re}(q_D)T + \frac{2q_1}{T} + \mathcal{O}(T^{-3}) \tag{S41}$$

For every data set that we use to determine $\beta(T)$, we also determine $\alpha(T)$. We then fit $\mathrm{Re}(\alpha(T))$ at large $T$ to $q_0 + 2\mathrm{Re}(q_{\mathrm{D}})T + 2q_1/T$ where $q_0$ and $q_1$ are real fitting parameter (and $q_{\mathrm{D}} = -\int_0^1 ds\, \lambda_+(s)$ is known). The measured $\mathrm{Re}(\alpha(T))$ and the corresponding fits are shown in the top portions of Supplemental Figures 4a-c.

Rounding the best-fit value of $q_0/2\pi$ to the nearest integer then provides $(n_{\circlearrowright} + n_{\circlearrowleft})$. Since $(n_{\circlearrowright} + n_{\circlearrowleft})$ and $(n_{\circlearrowright} - n_{\circlearrowleft})$ have the same parity, we define $n = (n_{\circlearrowright} + n_{\circlearrowleft}) \bmod 2$. All values of the complex Berry phase $\phi_B$ shown here are equal to $\pi n$ plus the $T$-independent fit coefficient of $\beta(T)$.

### *4.2 Determining the large-T asymptote of the data*

For the power series expansions described above to be valid, the data used for fitting must be sufficiently far into the adiabatic regime (corresponding to large $T$). However, for most control loops $\mathcal{C}$, the membrane's motion decays. In practice this sets an upper limit to $T$ (beyond which the membrane's motion is indistinguishable from its thermal fluctuations) which we denote $T_{\mathrm{SNR}}$. The value of $T_{\mathrm{SNR}}$ depends upon $\mathcal{C}$, as the normal modes' decay rate depends upon the control tones' powers and detuning.

To address this, for each $\mathcal{C}$ we perform the asymptotic fits to $\alpha(T)$ and $\beta(T)$ for $T_{\mathrm{min}} < T < T_{\mathrm{SNR}}$ using one of three possible options for $T_{\mathrm{min}}$:

***Option (1):*** $T_{\mathrm{min}} = 8(|\lambda_+ - \lambda_-|)^{-1}$. Here, $(|\lambda_+ - \lambda_-|)^{-1}$ is a typical time scale associated with adiabaticity in non-Hermitian systems.

***Option (2):*** For those $\mathcal{C}$ in which Option (1) leaves insufficient data for the asymptotic fit, we instead use $T_{\mathrm{min}} = \frac{2}{3} T_{\mathrm{SNR}}$.

***Option (3):*** For those $\mathcal{C}$ in which Option (1) results in $T_{\mathrm{min}} < \frac{1}{3} T_{\mathrm{SNR}}$, we instead use $T_{\mathrm{min}} = \frac{1}{3} T_{\mathrm{SNR}}$.

Examples of these fits are shown in Supplemental Figures 4a-c. In Supplemental Figure 4a the fits use Option (3); in Supplemental Figure 4b the fits use Option (1), and in Supplemental Figure 4c they use Option (2).

**Supplemental Note 5 Calculating the geometric phase**

This Supplemental Note reviews the calculation of the geometric phase $\phi_B$, both for cases in which there are simple analytic expressions, and for which it must be evaluated numerically. We also describe an easily visualizable relationship between $\phi_B$ and the shape of the "simple" control loops pictured in Fig. 2b.

***5.1 Geometric phase for "simple" control loops***

The eigenvalues of the traceless part of $H$ (see Eq S13) are $\lambda_\pm = \pm\sqrt{L^2 + MN}$, and the corresponding right (left) eigenvectors $\vec{\psi}_\pm(\vec{\xi}_\pm)$ are, up to an overall complex scaling factor, given by

$$\vec{\psi}_\pm = \begin{pmatrix} Me^{i\theta_{12}} \\ \pm\lambda - L \end{pmatrix}, \tag{S42}$$

$$\vec{\xi}_\pm = \pm\frac{1}{2\lambda M e^{i\theta_{12}}}\begin{pmatrix} \pm\lambda + L \\ Me^{i\theta_{12}} \end{pmatrix}. \tag{S43}$$

Note that this choice of eigenvectors satisfies the normalization condition $\vec{\xi}_i \cdot \vec{\psi}_j = \delta_{ij}$ (the dot product between two vectors is defined as $\vec{v} \cdot \vec{w} = v_1 w_1 + v_2 w_2$). For the "simple" control loops considered in the main text (which correspond to Fig. 2b), we only need consider the Berry connection along the $\theta_{12}$ direction of parameter space, which may be readily computed [6,7]:

$$\mathcal{A}_{\pm,\theta_{12}} = i\vec{\xi}_\pm \cdot \partial_{\theta_{12}}\vec{\psi}_\pm = -\frac{1}{2}(1 \pm L/\lambda) \tag{S44}$$

and the geometric phase for a circuit in which we ramp $\theta_{12}: 0 \rightarrow x$ is simply given by

$$\phi_{B,\pm} = \int_0^x d\theta_{12}\mathcal{A}_{\pm\theta_{12}} = -\frac{x}{2}(1 \pm L/\lambda) \tag{S45}$$

i.e. the accumulated geometric phase is proportional to the integrated change in the beatnote phase. For a complete phase winding $\theta_{12}: 0 \rightarrow 2\pi$, we have

$$\phi_{B,\pm} = -\pi(1 \pm L/\lambda) = \pm\pi(1 - L/\lambda) \tag{S46}$$

where in the last equality we have used the fact that $\phi_{B,\pm}$ is defined mod $2\pi$. Note that here, the eigenvector $\vec{\psi}_+$ need not refer to the least-damped eigenmode. Indeed, we have made no choice of the principal branch of $\lambda$ in the derivation above. In practice, we choose $+\lambda$ to be the eigenvalue with the largest imaginary component, which draws analogy to the Hermitian case in the limit $\mathrm{Im}(\lambda) \to 0$. For all data presented in this manuscript, this also fixes $+\lambda$ as the eigenvalue of the least-damped mode, and so for notational simplicity we define $\phi_B \equiv \phi_{B,+}$.

### *5.2 Geometric intuition for the complex Berry phase*

The parameterization of $H$ in terms of $L, M, N, \theta_{12}$ is convenient for computing the geometric phase accumulated in most of the experiments presented here, and in particular for the "simple" control loops illustrated in Fig. 2b. However, when considering more general loops it can be convenient to parameterize $H$ (up to its trace) in complex spherical coordinates (*8*)

$$H = \lambda \begin{pmatrix} \cos(\upsilon) & \sin(\upsilon)\, e^{-i\theta} \\ \sin(\upsilon)\, e^{i\theta} & -\cos(\upsilon) \end{pmatrix} = \vec{B} \cdot \vec{\sigma} \tag{S47}$$

$$\vec{B} = \lambda \begin{pmatrix} \sin(\upsilon)\cos(\theta) \\ \sin(\upsilon)\sin(\theta) \\ \cos(\upsilon) \end{pmatrix} \tag{S48}$$

with $\lambda, \theta, \upsilon \in \mathbb{C}$. In these coordinates, the eigenvalues are $\pm\lambda$, and the corresponding eigenvectors are given by

$$\vec{\psi}_+ = \begin{pmatrix} \cos(\upsilon/2) \\ \sin(\upsilon/2) e^{i\theta} \end{pmatrix}, \vec{\psi}_- = \begin{pmatrix} \sin(\upsilon/2) e^{-i\theta} \\ -\cos(\upsilon/2) \end{pmatrix},$$

$$\vec{\xi}_+ = \begin{pmatrix} \cos(\upsilon/2) \\ \sin(\upsilon/2)\, e^{-i\theta} \end{pmatrix}, \vec{\xi}_- = \begin{pmatrix} \sin(\upsilon/2)\, e^{i\theta} \\ -\cos(\upsilon/2) \end{pmatrix}. \tag{S49}$$

The Berry connection may be readily computed to be

$$\vec{\mathcal{A}}_\pm = i\vec{\xi}_\pm \cdot \nabla\vec{\psi}_\pm = \mp\frac{1}{2}(1-\cos(\upsilon))\hat{\theta}. \tag{S50}$$

Note that the Berry connection $\vec{\mathcal{A}}_\pm$ is a vector in Hamiltonian parameter space, and $\nabla$ is the gradient operator in this space. Thus, when considering a generic path in parameter space, we need only to consider the projection of the path to the complex $\theta$ plane.

For the loops considered in the main text, the experimental parameter $\theta_{12}$ maps onto $-\mathrm{Re}(\theta)$, meaning that only the real part of $\theta$ is varied by $2\pi$. In practice, the Hamiltonians accessible in these measurements have $\mathrm{Im}[\theta] \approx 0$, and thus for the remainder of this discussion we consider $\theta = -\theta_{12} \in \mathbb{R}$ (though the results remain unchanged for a closed circuit through arbitrary complex $\theta$). Thus, for the paths considered in the main text, we have (for the least-damped mode):

$$\begin{aligned}\phi_{\mathrm{B}} = \int_0^{2\pi} d\theta_{12}\mathcal{A}_{+,\theta_{12}} &= \pi\big(1-\cos(\upsilon)\big) \\ &= \pi\big(1 - \cos(\upsilon_{\mathrm{re}})\cosh(\upsilon_{\mathrm{im}}) + i\sin(\upsilon_{\mathrm{re}})\sinh(\upsilon_{\mathrm{im}})\big)\end{aligned} \tag{S51}$$

where we have written $\upsilon = \upsilon_{\mathrm{re}} + i\upsilon_{\mathrm{im}}$. It was noted in Ref. [7] that the line integral of $\vec{\mathcal{A}}_{\pm}$ over a closed circuit $\mathcal{C}$ may be converted, via Stokes' theorem, into the surface integral of a curvature over the area bounded by $\mathcal{C}$. Indeed, in complete analogy to the Hermitian Berry curvature, this curvature may be written (in $\mathbb{C}^3$) as the field strength tensor associated with a "magnetic field":

$$\vec{R}_{\pm} = \mp\frac{\vec{B}}{2B^3} \tag{S52}$$

where $\vec{B}$ is the complex vector defined above (see Ref. [8], Eq. 45). Therefore, the complex geometric phase can be interpreted as a flux enclosed by $\mathcal{C}$ in the presence of a "magnetic monopole" that is located at $\vec{B}\cdot\vec{B} = 0$ (i.e., where $H$ is degenerate).

In spite of this algebraic analogy between the Hermitian and non-Hermitian case, there are also important differences. In the Hermitian case, the space of $\vec{B}$ is $\mathbb{R}^3$, with the only degeneracy occurring at the point $\vec{B} = 0$. However, in the non-Hermitian case the space of $\vec{B}$ is six (real) dimensional and the degeneracy (i.e., $\vec{B}\cdot\vec{B} = 0$) occurs in a four-dimensional subspace, making the relationship between $\mathcal{C}$ and $\phi_{\mathrm{B}}$ challenging to visualize.

We may visualize the situation by decomposing

$$\vec{B}/\lambda = \cosh(\upsilon_{\mathrm{im}})\begin{pmatrix}\sin(\upsilon_{\mathrm{re}})\cos(\theta)\\ \sin(\upsilon_{\mathrm{re}})\sin(\theta)\\ \cos(\upsilon_{\mathrm{re}})\end{pmatrix} + i\sinh(\upsilon_{\mathrm{im}})\begin{pmatrix}\cos(\upsilon_{\mathrm{re}})\cos(\theta)\\ \cos(\upsilon_{\mathrm{re}})\sin(\theta)\\ -\sin(\upsilon_{\mathrm{re}})\end{pmatrix} = \vec{s}_{\mathrm{re}} + i\vec{s}_{\mathrm{im}} \tag{S53}$$

where $\vec{s}_{\mathrm{re}}$ and $\vec{s}_{\mathrm{im}}$ are vectors in $\mathbb{R}^3$. It may be readily verified that $\vec{s}_{\mathrm{re}}\cdot\vec{s}_{\mathrm{im}} = 0$. For loops of the type shown in Fig. 2b and in the limit $|\vec{B}_{\mathrm{im}}| \ll |\vec{B}_{\mathrm{re}}|$ (i.e. $\upsilon_{\mathrm{im}} \ll 1$) which is relevant for many of the measurements in this manuscript, the real component of the geometric phase

$$\mathrm{Re}(\phi_\mathrm{B}) = \pi\big(1 - \cos(\upsilon_\mathrm{re})\big) + \mathcal{O}({\upsilon_\mathrm{im}}^2) = \frac{\Upsilon}{2} + \mathcal{O}({\upsilon_\mathrm{im}}^2) \tag{S54}$$

is approximately equal to one-half the solid angle $\Upsilon$ subtended by the $\vec{s}_\mathrm{re}$ vector – a result familiar from the Hermitian case.

In addition, by noting that $|\vec{s}_\mathrm{re}| = \cosh(\upsilon_\mathrm{im}) = 1 + \mathcal{O}({\upsilon_\mathrm{im}}^2)$ and $|\vec{s}_\mathrm{im}| = |\sinh(\upsilon_\mathrm{im})| = |\upsilon_\mathrm{im}| + \mathcal{O}({\upsilon_\mathrm{im}}^3)$, we see that area of the mantle of the cone (see Supplemental Figure 3a) swept out by $\vec{s}_\mathrm{re}$ is $\widetilde{\Upsilon} = \pi\sin(\upsilon_\mathrm{re}) + \mathcal{O}({\upsilon_\mathrm{im}}^2)$, and thus

$$\mathrm{Im}(\phi_\mathrm{B}) = \pi\big(\sin(\upsilon_\mathrm{re})\sinh(\upsilon_\mathrm{im})\big) = \widetilde{\Upsilon}|\vec{s}_\mathrm{im}| + \mathcal{O}({\upsilon_{im}}^2). \tag{S55}$$

In the limit of $\upsilon \in \mathbb{R}$, (i.e. $|\vec{s}_\mathrm{im}| = 0$) this reduces to the standard (real) Hermitian Berry phase, even for Hamiltonians with complex $\lambda$.

### *5.3 Numerical method for calculating the geometric phase*

For "non-simple" control loops (i.e., not corresponding to Fig. 2b) there is no simple expression for $\phi_\mathrm{B}$, so we calculate it numerically as follows. First, we discretize the loop into $N$ equal steps, each starting at $s_k = k/N$ with $k = 0, 1, 2, \ldots, N$. In analogy to the Hermitian case (Eq. 4 in Ref. [9]), the non-Hermitian $\phi_\mathrm{B}$ can then be expressed as

$$\phi_{\mathrm{B},\pm} \approx -i\log\left(\prod_{k=0}^{N-1} \vec{\xi}_\pm(s_{k+1}) \cdot \overrightarrow{\psi}_\pm(s_k)\right). \tag{S56}$$

Thus, we calculate the geometric phase by computing the left and right eigenvectors at each step $s_k$ (while ensuring there is a consistent choice of gauge between all $N$ steps) and multiplying the chain of inner products above. For closed loops, this formulation is invariant to gauge choice, as $\overrightarrow{\psi}_\pm \to e^{i\alpha}\overrightarrow{\psi}_\pm$ and $\vec{\xi}_\pm \to e^{-i\alpha}\vec{\xi}_\pm$ for all $\alpha \in \mathbb{R}$. The calculations presented here use $N = 20{,}000$.

**Supplemental Note 6: Additional measurements**

This Supplemental Note presents additional measurements that use different types of control loops and different mechanical modes.

***6.1 Additional measurements using "simple" control loops***

In addition to the measurements presented in Fig. 3c (for which $\eta/2\pi = -50$), we have performed similar measurements over the $(P, \delta)$ plane for $\eta/2\pi = -90$ Hz and $\eta/2\pi = -20$ Hz. As in Fig. 3c, for each $P, \delta, \eta$, we measure $\beta(T)$ and $\alpha(T)$, and fit the data at large $T$ to extract the asymptote (i.e., the geometric phase $\phi_B$) as detailed in Supplemental Note 4.

Supplemental Figure 6a,b shows the geometric phases extracted from these measurements, along with the predicted values of $\phi_B$. We note that some loops in Supplemental Figure 6a have almost entirely real $\phi_B$, and that some loops in Supplemental Figure 6b have almost entirely imaginary $\phi_B$.

Supplemental Figure 7 also shows $\beta(T)$ as a function of $(P, \delta)$ for $\eta/2\pi = -20$ Hz, $-50$ Hz, and $-90$ Hz, similar to Fig. 3a,b. Note that for every measurement in which the fit to $\mathrm{Re}(\alpha(T))$ gives $n = 1$, we have shifted the entire data set by $\pi$ to reflect the different choice of branch (see Supplemental Note 4.1).

***6.2 Control loops with non-constant speed***

The data shown in Figs. 2,3 use loops in which $(P_1, P_2, \delta, \eta)$ are fixed while the beat note phase is varied so that $\theta_{12}(t) = \pm 2\pi t/T$. However, it is straightforward to realize other types of loops. In this section, we describe measurements using loops in which $(P_1, P_2, \delta, \eta)$ are fixed but the choice of $\theta_{12}(t)$ varies. This provides a demonstration of one of the key features of geometric phase: that it is determined entirely by the shape of $\mathcal{C}$, and does not depend on the time-dependence used to traverse $\mathcal{C}$. These measurements used the membrane's (3,3) and (5,3) modes, and $(P_1, P_2, \Delta/2\pi, \eta/2\pi) =$ (21 μW, 21 μW, –1.6 MHz, –27.5 Hz).

Listed below are five functional forms for $\theta_{12}(t)$ that vary monotonically from 0 to $\pm 2\pi$ as $t$ increases from 0 to $T$:

**(A)** $\theta_{12}(t) = \pm 2\pi (t/T)^2$

**(B)** $\theta_{12}(t) = \pm 2\pi \sqrt{t/T}$

**(C)** $\theta_{12}(t) = \pm 2\pi \sin(\pi t/2T)$

**(D)** $\theta_{12}(t) = \pm 2\pi (6(t/T)^5 - 15(t/T)^4 + 10(t/T)^3)$

**(E)** $\theta_{12}(t) = \pm 2\pi (t/T + \sin(\pi t/T)^2/4)$

These functions are plotted in the left column of the corresponding panels of Supplemental Figure 8a-e.

While $\phi_B$ is predicted to be independent of $\theta_{12}(t)$, the manner in which $\beta(T)$ approaches $\phi_B$ for large $T$ does depend on $\theta_{12}(t)$. The central and right columns of Supplemental Figure 8a-e demonstrate that $\beta(T)$ indeed changes with $\theta_{12}(t)$, but that for large $T$ it approaches the predicted $\phi_B$. This is shown explicitly in Supplemental Figure 8f, which plots the asymptotic values of $\beta(T)$ (extracted as in Supplemental Note 3) along with the predicted $\phi_B$.

### *6.3 "Non-simple" control loops*

In this section we describe measurements that use "non-simple" control loops, by which we mean loops that involve varying the control tones' powers and detunings (in addition to $\theta_{12}$).

For these measurements, we modify the setup shown in Supplemental Figure 1 to vary $\delta$ and $P_2$ (in addition to $\theta_{12}$) in real time. Control loops realized in this way can be parameterized as: $\vec{X}(s) = (P_1, P_2(s), \delta(s), \eta, \theta_{12}(s))$. Note that for all three of the $\vec{X}(s)$ described here, one mode has lower damping during the entire loop, so that the adiabatic theorem applies for large $T$.

The first set of non-simple loops that we study are of the form:

$$\vec{X}_1(s) = \left(P_1, P_{2,\text{start}} + P_a \sin(2\pi s), \frac{\delta_{\text{min}}+\delta_{\text{max}}}{2} + \frac{\delta_{\text{min}} - \delta_{\text{max}}}{2}\cos(2\pi s), \eta, 2\pi s\right) \quad \text{(S57)}$$

where $s = \pm t/T$, $t \in [0, T]$, and the loop parameters are: $P_1 = 15$ μW, $\eta/2\pi = -50$ Hz, $P_{2,\text{start}} = 15$ μW, $P_a = 5$ μW, and $\delta_{\text{min}}/2\pi = -1.5$ MHz.

Loops with $\delta_{\text{max}}/2\pi = -1.5, -1.25, -1.0, -0.75, -0.5, -0.25$ MHz are illustrated in Supplemental Figure 9a. Measurements using these loops were carried out in the manner described in the main text, i.e., traversing each loop both "forward" and "backward" (corresponding to the choice of sign for $s(t)$) and taking the difference of the accumulated phases $\beta$ for a range of $T$. These measurements were made using the membrane modes (3,3) and (5,2), and are shown in Supplemental Figure 9b.

The second set of non-simple loops are of the form:

$$\vec{X}_2(s) = \left(P_1, P_{2,\text{start}} + P_a \sin(2\pi s), \frac{\delta_{\text{min}}+\delta_{\text{max}}}{2} + \frac{\delta_{\text{min}} - \delta_{\text{max}}}{2} \cos(2\pi s), \eta, \theta_{12,\text{max}} \sin^2(\pi s)\right) \quad \text{(S58)}$$

where $s = \pm t/T$, $t \in [0, T]$, and the loop parameters are: $P_1 = 15$ μW, $\eta/2\pi = -50$ Hz, $P_{2,\text{start}} = 15$ μW, $P_a = 10$ μW, $\delta_{\text{min}}/2\pi = -1.5$ MHz, and $\delta_{\text{max}}/2\pi = -0.25$ MHz.

These loops differ qualitatively from the ones described by $\vec{X}_1$ in that $\theta_{12}$ is not "wrapped" by $2\pi$ in completing the loop. Loops with $\theta_{12,\text{max}} = (0, 2\pi/5, 4\pi/5, 6\pi/5, 8\pi/5, 2\pi)$ are illustrated in Supplemental Figure 9c. Measurements using these loops were carried out in the same fashion (and using the same modes) as for the $\vec{X}_1$ loops, and are shown in Supplemental

Figure 9d.

The third set of non-simple loops are of the form:

$$\vec{X}_3(s) = \left(P_1, P_{2,\text{start}} + P_{2,a}\left|\sin(\pi s) + \frac{1}{4}\sin^2(2\pi s)\right|, \delta, \eta, \theta_{12,\text{a}} \sin\left[2\pi s + \frac{1}{2}\sin(2\pi s)\right]\right) \quad \text{(S59)}$$

where $s = \pm t/T$, $t \in [0, T]$, and the loop parameters are: $P_1 = 21$ μW, $P_{2,\text{start}} = 5$ μW, $P_{2,a} = 22$ μW, $\delta/2\pi = -0.95$ MHz, and $\eta/2\pi = -30$ Hz. This is the form of the loop used as $\mathcal{C}_{\text{amp}}$ in the demonstration of SSGG (Fig 5), and also uses the third control tone shown in Fig. 4a. Loops with $\theta_{12,\text{a}} = (\pi/6, \pi/3, \pi/2, 2\pi/3, 5\pi/6, \pi)$ are illustrated in Supplemental Figure 9e. Measurements using these loops carried out in the same fashion as for the $\vec{X}_1$ loops (but using the membrane modes (3,3) and (5,3)) are shown in Supplemental Figure 9f.

### *6.4 Measurements using a different mechanical mode*

The measurements presented in Figs. 2,3 (and Supplemental Figure 3; 4; 6a,b; 7; 9b,d; 10a-c) used the membrane's (3,3) and (5,2) normal modes. In contrast the measurements in Fig. 4 (and Supplemental Figures 7, 8, 9f, 6c-e, 10d-g) used the (3,3) and (5,3) normal modes.

To illustrate that the phenomena studied here are generic (rather than specific to a particular pair of modes), here we describe measurements that are analogous to those in Figs. 2 – 4, but which use the (3,3) and (5,3) normal modes.

The parameters of the (5,3) mode are given in Table S1. To couple the (3,3) and (5,3) modes, we redefine the control tones' common and differential detunings: $\Delta_1 = -\omega_1^{(0)} + \delta$ and $\Delta_2 = -\omega_3^{(0)} + \delta + \eta$ such that $\left(\omega_3^{(0)} - \omega_1^{(0)}\right) - (\Delta_1 - \Delta_2) = \eta$.

Supplemental Figure 6c shows measurements of these modes' discriminant $D = (\lambda_+ - \lambda_-)^2$ as a function of $(P, \delta)$. As in Fig. 1d, the eigenvalues $\lambda_\pm$ of this coupled system are determined from measurements of the mechanical susceptibility. Also shown in Supplemental Figure 6c is a fit to $D(\delta, P)$ as described in Supplemental Note 2.

The left-hand column of Supplemental Figure 6d shows measurements of $\phi_\text{B}$ for "simple" control loops (i.e., with $\theta_{12}(t) = \pm 2\pi t/T$) for values of $(P, \delta)$ lying in the dashed rectangle of Supplemental Figure 6c. The right-hand column shows the calculated values of $\phi_\text{B}$ for each of these loops.

Supplemental Figure 6e shows parametric plots of $\beta(T)$ for each of these loops. The solid star shows the value of $\phi_\text{B}$ determined by fitting $\beta(T)$ for large $T$ (as described in Supplemental Note 4), and the hollow star shows the calculated value of $\phi_\text{B}$.

Supplemental Figure 10 shows measurements of $\text{Im}(\phi_\text{B})$ for control paths that are not closed loops.

**Supplemental Note 7: Berry phase on open paths**

The phase of an oscillator is only defined relative to a reference or gauge, which may vary within the space of control parameters. As a result, the (real-valued) phase accumulated during a control path $\mathcal{C}$ has a gauge-invariant definition only if $\mathcal{C}$ is a loop. In this section, we show that the imaginary part of the Berry connection is gauge invariant, from which it follows that the imaginary part of the geometric phase is also gauge independent, regardless of whether or not the path is closed. This discussion is similar to that of Ref. [6]. Qualitatively, this may be viewed as a reflection of the fact that an oscillator's amplitude does not depend on a gauge choice [10].

To begin, consider a choice of right and left eigenvectors $\vec{\psi}_{1,2}, \vec{\xi}_{1,2}$ that obey the joint normalization condition:

$$\vec{\xi}_{\mathrm{i}} \cdot \vec{\psi}_{\mathrm{j}} = \delta_{ij} \tag{S60}$$

An arbitrary state $\vec{d}$ may be decomposed as $\vec{d} = d_1\vec{\psi}_1 + d_2\vec{\psi}_2$ with complex coefficients $d_{1,2} = \vec{\xi}_{1,2} \cdot \vec{d}$. In this case, a state $\vec{d}$ that is proportional to the eigenmode $\vec{\psi}_i$ has population $|d_i|^2$.

Now, consider a different choice of right eigenvectors $\vec{\psi'}_{1,2} = x_{1,2}\vec{\psi}_{1,2}$ with nonzero constants $x_{1,2}$. To satisfy the normalization condition above, the corresponding left eigenvectors must be $\vec{\xi'}_{1,2} = \vec{\xi}_{1,2}/x_{1,2}$. However, upon making this choice, the population of $\vec{d}$ in eigenmode $\vec{\psi'}_i$ is

$$|d_i'|^2 = \left|\vec{\xi'}_i \cdot \vec{d}\right|^2 = \left|\frac{\vec{\xi}_{\mathrm{i}}}{x_i} \cdot \vec{d}\right|^2 = |d_i|^2/|x_i|^2 \tag{S61}$$

i.e. the population depends explicitly on $|x_i|$. Since the population of a mode encodes an absolute physical quantity (in this case, it is proportional to the energy stored in the oscillator), it must be independent of the choice of gauge (i.e. the choice of $\vec{\psi}_{1,2}, \vec{\xi}_{1,2}$). This is ensured by imposing an additional normalization condition:

$$\left|\vec{\psi}^*_{1,2} \cdot \vec{\psi}_{1,2}\right|^2 = 1 \tag{S62}$$

Thus, any choice of right eigenvectors (i.e. choice of gauge) obeys $\vec{\psi'}_{1,2} = x_{1,2}\vec{\psi}_{1,2}$, with $|x_i| = 1$. Put another way, to preserve the gauge independence of populations, the only allowed gauge transformations of right eigenmodes $\vec{\psi}_{1,2}$ are unitary complex numbers, just as in the Hermitian case.

Under such gauge transformations, the real part of the Berry connection $\mathcal{A}_i = \vec{\xi}_i(\mathrm{s}) \cdot$

$i\,\partial_s \vec{\psi}_i(s)$ changes, but the imaginary part does not. It follows that the imaginary component of the geometric phase is gauge invariant for any path (including open ones). When calculating the eigenvector amplitudes $\bar{c}_i$ from ringdown measurements (at both the beginning and the end of the circuit), the normalized eigenvectors that we use to construct $S_{\mathrm{D}}$ (Eq. S19) satisfies this restriction.

Supplemental Figure 10 shows measurements of $\mathrm{Im}(\phi_{\mathrm{B}})$ for a family of paths, some of which are not loops. They are realized by fixing $P$ and $\delta$, and ramping $\theta_{12}$ from 0 to $\pm(N/5)2\pi t/T$ for $N = 0,1,2,3,4,5$ (Supplemental Figure 10a), resulting in non-loop paths for $1 \leq N \leq 4$. Supplemental Figure 10b shows $\mathrm{Im}(\beta(T))$ for each $N$ (data and theory) and Supplemental Figure 10c shows $\mathrm{Im}(\phi_{\mathrm{B}})$ determined by fitting $\beta(T)$ as described in Supplemental Note 3.

## Supplemental Note 8: Steady State Geometric Gain (SSGG)

This Supplemental Note presents a detailed discussion of SSGG, and the range of control loops that can be used to produce it.

### *8.1 Defining SSGG*

Figure 4 of the main text shows that a collection of linear, lossy elements can produce continuous gain if the elements' parameters are varied slowly. While the flow of energy into or out of a system with a time-dependent Hamiltonian can be conceptualized in various ways, the behavior described here is conveniently described in terms of the complex geometric phase. We refer to this mechanism as steady-state geometric gain (SSGG). Here we provide a concrete, platform-independent definition of SSGG.

To focus the discussion, we consider systems that consist of two harmonic modes whose parameters are repeatedly tuned around a control loop $\mathcal{C}_{\mathrm{amp}}$, with each traversal using the same time dependence $s(t/T)$ and the same duration $T$. It is straightforward to generalize the concept of SSGG to a larger number of modes and to more complicated control paths.

We take the following four conditions as defining SSGG:

(1) There is a mode that is the least-damped mode for all of $\mathcal{C}_{\mathrm{amp}}$.

(2) When the system is prepared in this least-damped mode, its state after a single traversal of $\mathcal{C}_{\mathrm{amp}}$ using $s(t/T)$ with duration $T$ is well-approximated as the initial state multiplied by $e^{-i\phi(T)}$ where the complex phase $\phi(T)$ is given by Eq. 2.

(3) For each such traversal, this mode's dynamical phase contributes loss: $\mathrm{Im}(\phi_{\mathrm{D}}) < 0$.

(4) For each such traversal, this mode's total gain $\mathrm{Im}[\phi(T)]$ lies in $\Gamma$, the pink region of Supplemental Figure 5c.

When Condition (1) is met, the adiabatic theorem [5] guarantees that Condition (2) is also met for sufficiently large $T$. We take $T > T_{\mathrm{ad}}$ as the threshold for this condition, where $T_{\mathrm{ad}} = \max_{\mathcal{C}}(|\lambda_{+}(s) - \lambda_{-}(s)|^{-1})$.

Condition (3) requires that the system's dynamical phase contributes loss. This condition is included because otherwise the system can serve as steady-state amplifier without the geometric phase (i.e., just by using large $T$ and relying on the dynamical phase).

Condition (4) identifies the features that $\mathrm{Im}(\phi(T))$ must exhibit (upon a single traversal of $\mathcal{C}_{\mathrm{amp}}$ using time dependence $s$ and duration $T$) for the system's gain to be attributed to the geometric phase. These are:

(4a) $T > T_{\mathrm{ad}}$

(4b) $\mathrm{Im}[\phi(T)] > 0$

(4c) $\mathrm{Im}[\phi(T)] < -\mathrm{Im}(\phi_\mathrm{B})$

(4d) $\mathrm{Im}[\phi(T)] < \mathrm{Im}(\phi_\mathrm{D}) - 2\mathrm{Im}(\phi_\mathrm{B})$

Each of the Conditions (4a – 4d) has a straightforward physical interpretation.

Condition (4a) ensures that the system's dynamics are compatible with the notion of adiabaticity, and hence are described by Eq. 2.

Condition (4b) ensures that the mode experiences net gain upon each traversal of $\mathcal{C}_\mathrm{amp}$.

Conditions (4c) and (4d) ensure that the net gain can be attributed to the geometric gain. Specifically, Condition (4c) requires that the net gain would be negative without the contribution from the geometric gain. Condition (4d) requires that the geometric gain accounts for the majority of the difference between the dynamical loss $\mathrm{Im}(\phi_\mathrm{D})$ and the net gain; this ensures that the net gain is not attributable to the higher-order terms [i.e., $\mathcal{O}(T^{-1})$] in Eq. 2.

For a given $\mathcal{C}_\mathrm{amp}$ and $s$, Conditions (4a – 4d) can be visualized by noting that they each bisect the plane spanned by $\mathrm{Im}[\phi(T)]$ and $T$ (Supplemental Figure 5c, grey lines). Together, they define a region (Γ) with the following interpretation: for this $\mathcal{C}_\mathrm{amp}$ and $s(t/T)$, SSGG results if and only if $T$ is chosen so that $\mathrm{Im}[\phi(T)]$ lies within Γ.

## 8.2 Identifying control loops that produce SSGG

Conditions (1 – 3) depend only on the "static" or "instantaneous" properties of the system (e.g., its eigenvalues at each value of $s$). As a result, for a given physical system and control loop, it is relatively straightforward to determine whether they are satisfied.

In contrast, Condition (4) depends on the system's dynamics as governed by Eq. 1. For a specific $\mathcal{C}_\mathrm{amp}$ and $s$, numerical integration of Eq. 1 (for various choices of $T$) can be used to determine whether or not $\mathrm{Im}[\phi(T)]$ passes through Γ. In practice, this approach was used to identify the $\mathcal{C}_\mathrm{amp}$ used in Fig. 4.

In addition to explicit numerical integration, it would be helpful to also have a simple criteria that could be used to identify which $\mathcal{C}_\mathrm{amp}$ and $s$ can produce SSGG. As a practical matter, this would facilitate the design of SSGG devices. More abstractly, it could also provide a useful measure of how common such loops are. We are not aware of any simple criteria that determine whether an arbitrary $\mathcal{C}_\mathrm{amp}$ and $s$ can meet Condition (4). Instead, we consider specific families of loops that are parameterized by a few variables, and determine the range of these variables for which the corresponding loops satisfy Condition (4). Below, we consider two such families, and find that in both cases a wide range of loops satisfy Condition (4).

***SSGG with simple loops:*** The first loops we consider are the simple ones (Fig. 2b), which can be parameterized as:

$$H(t) = \begin{pmatrix} \mathcal{T} + L & Me^{-2\pi it/T} \\ Ne^{2\pi it/T} & \mathcal{T} - L \end{pmatrix} \tag{S63}$$

where $\mathcal{T}, M, N$, and $L$ are complex constants. For $H(t)$ of this form, the eigenvalues are time-independent and given by $\lambda_{\pm} = \mathcal{T} \pm \lambda_0$, where $\lambda_0 = \sqrt{L^2 - MN}$. Since we are interested in the less-damped mode, we take $\mathrm{Im}(\lambda_0) \geq 0$. In this discussion, the trace $\mathcal{T}$ plays an important role, as its imaginary part contributes to the system's loss (in particular, to $\mathrm{Im}(\phi_{\mathrm{D}})$).

Equation 1 with $H(t)$ can be solved analytically to give

$$\phi(T) = \mathcal{T}T + \sqrt{\lambda_0^2 T^2 - 2\pi LT + \pi^2} - \pi \tag{S64}$$

It is straightforward to show that in the $T \to \infty$ limit, $\phi(T) \to \phi_{\mathrm{D}} - \phi_{\mathrm{B}}$, where $\phi_{\mathrm{D}} = q_{\mathrm{D}}T = (\mathcal{T} + \lambda_0)T$ and $\phi_{\mathrm{B}} = \pi\left(\frac{L}{\lambda_0} + 1\right)$.

These loops always meet Condition (1), and always meet Condition (2) in the large-$T$ limit. Meeting Condition (3) amounts to the simple constraint $\mathrm{Im}(\mathcal{T} + \lambda_0) < 0$. As a result, identifying the loops that can produce SSGG amounts to finding the values of $\mathcal{T}$, $\lambda_0$, and $L$ that meet Condition (4).

Condition (4) requires that the function $\mathrm{Im}\big(\phi(T)\big)$ passes through Γ, which in turn requires that $\mathrm{Im}\big(\phi(T)\big)$ intersects[1] at least one of the boundaries of Γ. Here we derive an analytic expression for the loops that intersect the lower boundary of Γ. Similar expressions can be derived for the other boundaries of Γ.

$\mathrm{Im}\big(\phi(T)\big)$ intersects the lower boundary of Γ if there is a loop duration $T_*$ such that the following two conditions hold:

(*A*) $T_{\mathrm{ad}} < T_* < T_{\mathrm{c}}$

(*B*) $\mathrm{Im}\big(\phi(T_*)\big) = 0$

where $T_{\mathrm{c}} = 2\mathrm{Im}(\phi_{\mathrm{B}})/\mathrm{Im}(q_{\mathrm{D}})$.

Considering (*B*) first, we rewrite it as

$$\mathrm{Im}\big(\phi(T_*)\big) = 0 \tag{S65}$$

[1] Assuming this intersection is non-tangential.

$$\mathrm{Im}\left(\mathcal{T}T_* + \sqrt{\lambda_0^2 T_*^2 - 2\pi L T_* + \pi^2} - \pi\right) = 0 \quad \text{(S66)}$$

$$\mathrm{Im}\left(\sqrt{\lambda_0^2 T_*^2 - 2\pi L T_* + \pi^2}\right) = \mathrm{Im}(-\mathcal{T}T_*) \quad \text{(S67)}$$

$$\sqrt{\lambda_0^2 T_*^2 - 2\pi L T_* + \pi^2} = -\mathcal{T}T_* + r \quad \text{(S68)}$$

where $r$ is a real number. This is equivalent to:

$$\lambda_0^2 T_*^2 - 2\pi L T_* + \pi^2 = \mathcal{T}^2 T_*^2 - 2\mathcal{T}T_* r + r^2 \quad \text{(S69)}$$

The imaginary part of Eq. S69 is

$$T_*^2 \mathrm{Im}(\lambda_0^2 - \mathcal{T}^2) - 2\pi T_* \mathrm{Im}(L) + 2T_* r \mathrm{Im}(\mathcal{T}) = 0 \quad \text{(S70)}$$

For any value of $r$, one solution of this equation is $T_* = 0$ (which is also obviously a solution of Eq. S66). For $T_* \neq 0$ and $\mathrm{Im}(\mathcal{T}) \neq 0$ we have

$$r = \frac{T_*^2 \mathrm{Im}(\mathcal{T}^2 - \lambda_0^2) + 2\pi T_* \mathrm{Im}(L)}{2T_* \mathrm{Im}(\mathcal{T})} = \frac{T_*(\mathcal{T}_\mathrm{r}\mathcal{T}_\mathrm{i} - \lambda_\mathrm{r}\lambda_\mathrm{i}) + \pi L_\mathrm{i}}{\mathcal{T}_\mathrm{i}} \quad \text{(S71)}$$

where $\mathcal{T} = \mathcal{T}_\mathrm{r} + i\mathcal{T}_\mathrm{i}$, $\lambda_0 = \lambda_\mathrm{r} + i\lambda_\mathrm{i}$, and $L = L_\mathrm{r} + iL_\mathrm{i}$. Inserting this into the real part of Eq. S69 gives:

$$T_*^2 \mathrm{Re}(\lambda_0^2 - \mathcal{T}^2) - 2\pi T_* \mathrm{Re}(L) + 2T_* r \mathrm{Re}(\mathcal{T}) + \pi^2 - r^2 = 0 \quad \text{(S72)}$$

$$\frac{(\mathcal{T}_\mathrm{i}^2 - \lambda_\mathrm{i}^2)(\mathcal{T}_\mathrm{i}^2 + \lambda_\mathrm{r}^2)}{\mathcal{T}_\mathrm{i}^2} T_*^2 + 2\pi\left(\frac{L_\mathrm{i}\lambda_\mathrm{r}\lambda_\mathrm{i}}{\mathcal{T}_\mathrm{i}^2} - L_\mathrm{r}\right) T_* + \pi^2\left(1 - \frac{L_\mathrm{i}^2}{\mathcal{T}_\mathrm{i}^2}\right) = 0 \quad \text{(S73)}$$

$$(\mathcal{T}_\mathrm{i}^2 - \lambda_\mathrm{i}^2)(\mathcal{T}_\mathrm{i}^2 + \lambda_\mathrm{r}^2) T_*^2 + 2\pi(L_\mathrm{i}\lambda_\mathrm{r}\lambda_\mathrm{i} - \mathcal{T}_\mathrm{i}^2 L_\mathrm{r}) T_* + \pi^2(\mathcal{T}_\mathrm{i}^2 - L_\mathrm{i}^2) = 0 \quad \text{(S74)}$$

This is a quadratic equation for $T_*$ which we write as

$$aT_*^2 + bT_* + c = 0 \tag{S75}$$

with $a, b, c$ all real numbers. Its two solutions

$$T_* = T_\pm = \frac{-b \pm \sqrt{b^2 - 4ac}}{2a} \tag{S76}$$

are real for $b^2 > 4ac$, a condition that corresponds to

$$(L_\mathrm{i}\lambda_\mathrm{r}\lambda_\mathrm{i} - \mathcal{T}_\mathrm{i}^{\,2}L_\mathrm{r})^2 > (\mathcal{T}_\mathrm{i}^{\,2} - \lambda_\mathrm{i}^2)(\mathcal{T}_\mathrm{i}^{\,2} + \lambda_\mathrm{r}^2)(\mathcal{T}_\mathrm{i}^{\,2} - L_\mathrm{i}^2) \tag{S77}$$

Thus, ($B$) is satisfied for $T_* = T_\pm$ (together with the constraint given by Eq. S77). To produce SSGG, a loop must also satisfy ($A$), which amounts to the requirement that either $T_\mathrm{ad} < T_+ < T_\mathrm{c}$ or $T_\mathrm{ad} < T_- < T_\mathrm{c}$ . These two conditions define the range of the five parameters $\mathcal{T}_\mathrm{i}, \lambda_\mathrm{r}, \lambda_\mathrm{i}, L_\mathrm{r}, L_\mathrm{i}$ over which SSGG occurs (note that $\mathcal{T}_\mathrm{r}$ does not appear in any of these expressions).

This range is shown in Supplemental Figure 5a, from which it is clear that SSGG can be produced by a wide range of simple loops. As mentioned above, this region represents only a subset of the loops that realize SSGG, as it does not include loops for which $\mathrm{Im}[\phi(T)]$ does not intersect the bottom boundary of $\Gamma$, but does intersect the other boundaries. These additional loops can be identified by repeating the same approach as above. This results in a quartic polynomial equation for $T_*$ (rather than quadratic as in Eq. S74), and hence more cumbersome expressions. Nevertheless, it would be straightforward to add these solutions to the region shown in Supplemental Figure 5a.

***SSGG with a class of non-simple loops:*** For a family of control loops without an analytic solution, the range of parameters satisfying Condition (4) can be identified by solving Eq. 1 numerically. Here, we consider a family of loops that may be regarded as representative of smooth (but non-simple) loops. They are qualitatively similar to the loop used in Fig. 4, and are given by

$$H(s) = \vec{B}(s) \cdot \vec{\sigma} + \mathcal{T}\mathbb{I}_2 \tag{S78}$$

with

$$B_x(s) = a\left[4 + \cos(2\pi s) + \cos(4\pi s) - i\left(4 + \cos(2\pi s) + \frac{4}{5}\cos(4\pi s) + \frac{3}{10}\cos(8\pi s)\right)\right]$$

$$B_y(s) = b\left[2\sin(2\pi s) + \frac{3i}{2}\left(\frac{1}{5} + \sin(2\pi s)\right)\right]$$

$$B_z(s) = c\left[4 + \frac{1}{6}\left(\sin(2\pi s) - \cos(4\pi s)\right) + i\left(2 + \frac{1}{7}\sin(2\pi s) - \frac{1}{7}\cos(4\pi s)\right)\right]$$

$$\mathcal{T} = 20 - 5i \tag{S79}$$

Here $a, b, c$ are the parameters used to vary the loop's shape. An example of one such loop is shown in Supplemental Figure 5b.

It is straightforward to solve Eq. 1 numerically for loops with various $a, b, c$. A typical example of $\mathrm{Im}[\phi(T)]$ resulting from one such solution is shown in Supplemental Figure 5c. In this particular case, $\mathrm{Im}[\phi(T)]$ lies within the region Γ for $0.047 < T < 0.41$.

Supplemental Figure 5d,e shows the result of performing this analysis for values of $(a, b, c)$ ranging from $0.05$ to 3, in increments of $0.05$. Each value of $(a, b, c)$ that satisfies Conditions $(1 - 4)$ (and hence, for which $\mathrm{Im}[\phi(T)]$ passes through Γ) is represented by an orange block in the figure. As with the simple loops considered in §8.3, SSGG occurs over a substantial range of the loop parameters.

| Parameter | Best fit value | Drift range |
|---|---|---|
| $\omega_1^{(0)}/2\pi$ | 2.423969 MHz | 20 Hz |
| $\omega_2^{(0)}/2\pi$ | 3.076488 MHz | 20 Hz |
| $\omega_3^{(0)}/2\pi$ | 3.331064 MHz | 20 Hz |
| $\gamma_1^{(0)}/2\pi$ | 3.6 Hz | 0.5 Hz |
| $\gamma_2^{(0)}/2\pi$ | 16.2 Hz | 0.5 Hz |
| $\gamma_3^{(0)}/2\pi$ | 3.35 Hz | 0.5 Hz |
| $\kappa/2\pi$ | 2.32 MHz | 0.1 MHz |
| $g_1/2\pi$ | 4.4 Hz | 0.1 Hz |
| $g_2/2\pi$ | 3.9 Hz | 0.1 Hz |
| $g_3/2\pi$ | 2.64 Hz | 0.1 Hz |
| $A_1/2\pi$ | –2.8 μHz | 0.5 μHz |
| $A_2/2\pi$ | –4.0 μHz | 0.5 μHz |
| $A_3/2\pi$ | –3.58 μHz | 0.5 μHz |
| $\Delta_{\mathrm{off}}/2\pi$ | –10 kHz | 10 kHz |

**Supplemental Table 1:** The best-fit values of the parameters appearing in the optomechanical Hamiltonian. For all quantities, the subscript denotes the mechanical mode: $1 \rightarrow (3,3)$, $2 \rightarrow (5,2)$, $3 \rightarrow (5,3)$. The best-fit values are given for a typical day. Small drifts in these parameters are tracked throughout the measurements prescribed here, and are included in all comparisons with theory. The rightmost column gives the extent of the parameters' drift over several months.

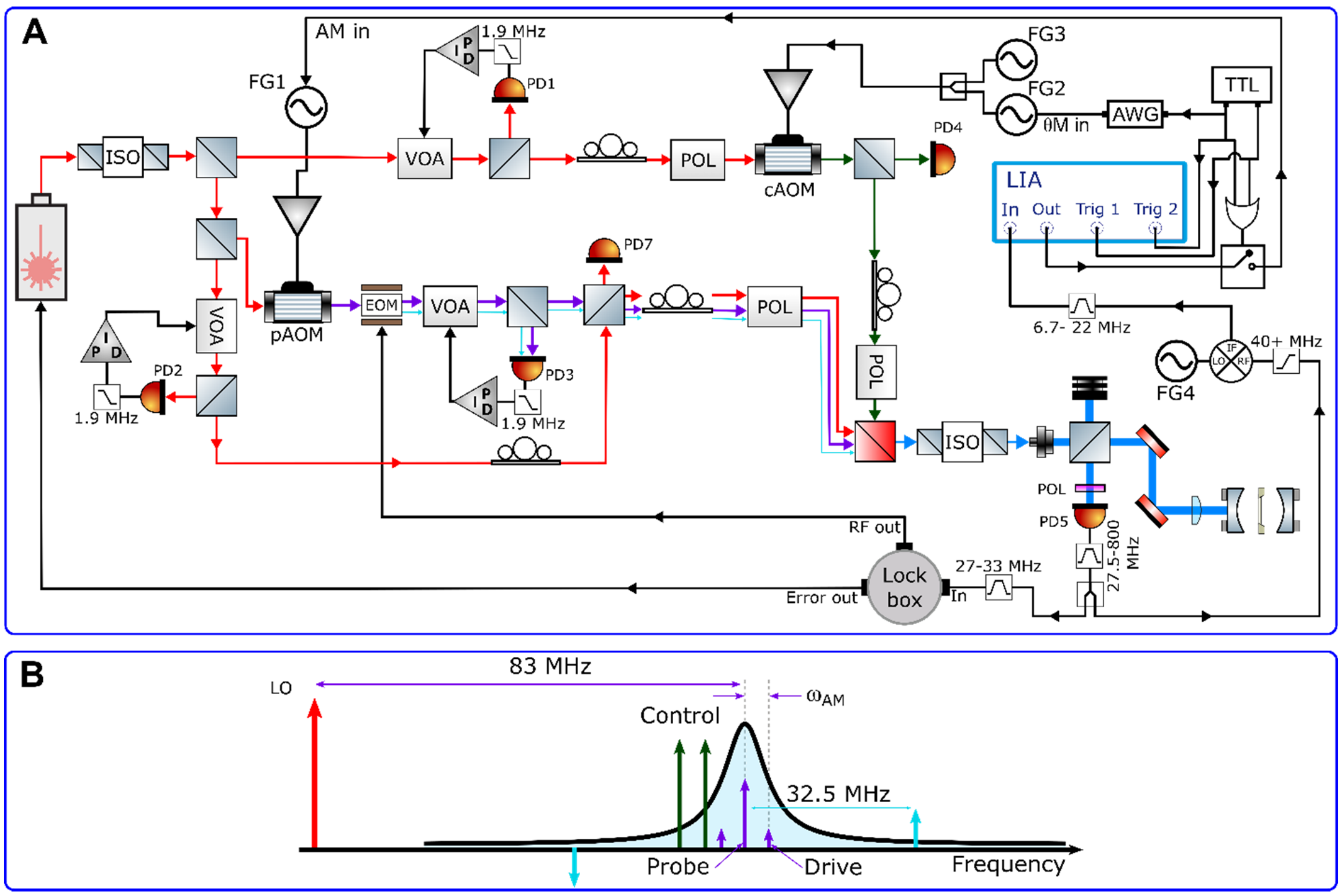

**Supplemental Figure 1: (a)** Detailed experimental setup. ISO: isolator. VOA: variable optical attenuator. POL: polarizer. AOM: acousto-optic modulator. EOM: electro-optic modulator. PD: photodiode. FG: function generator. AWG: arbitrary waveform generator. LIA: lock in amplifier. Black lines: electronic paths. Colored lines: optical paths. **(b)** Frequency space diagram showing all the optical tones (arrows) and the optical cavity resonance (black curve). Red: local oscillator. Dark green: control tones. Purple: probe beam and its AM sidebands (used to drive the membrane). Blue: the probe beam's FM sidebands, used for the PDH lock.

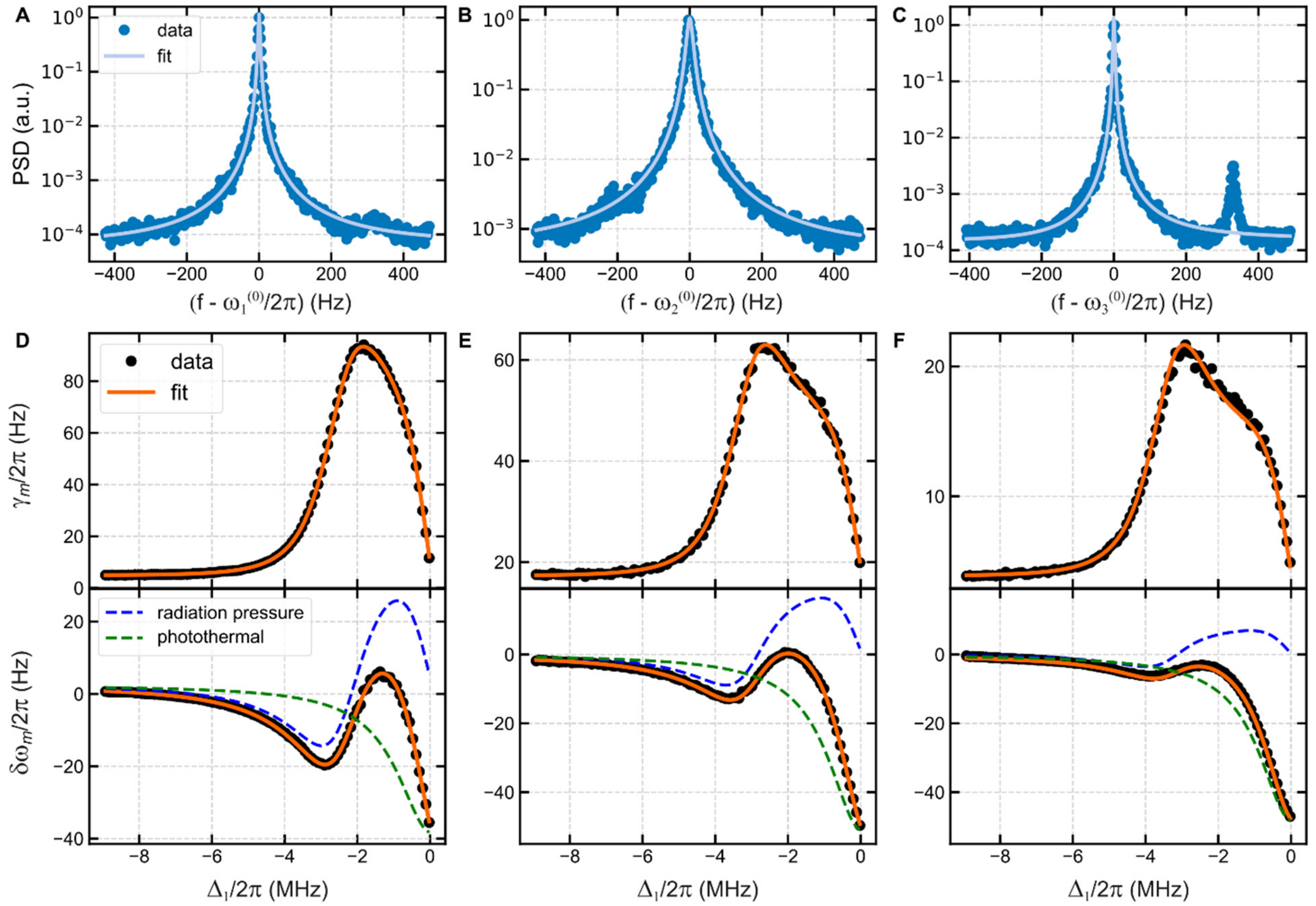

**Supplemental Figure 2:** Characterizing the membrane's modes and their optomechanical coupling to the cavity. (**a** – **c**) Representative spectra of the membrane's undriven motion at frequencies near the resonance of each of the three modes used in this work. (a) The membrane's (3,3) mode: $\omega_1^{(0)}/2\pi = 2.423969$ MHz. (b) The membrane's (5,2) mode: $\omega_2^{(0)}/2\pi = 3.076488$ MHz . (c) The membrane's (5,3) mode: $\omega_3^{(0)}/2\pi = 3.331064$ MHz. In (c), the small peak near 325 Hz is an aliasing artifact due to the motion of the (3,5) mode. **(d – f)** The dynamical back-action in the presence of a single laser tone with $P_1 = 17\ \mu$W and detuning $\Delta_1$ (relative to the optical cavity resonance) for the (3,3) mode in (d), the (5,2) mode in (e), and the (5,3) mode in (f). Upper (lower) panels: the shift in the mode's damping (frequency). Orange curve: fit to the optomechanical model. This model includes only radiation pressure for $\delta\gamma_i$, while for $\delta\omega_i$ it includes both radiation pressure (blue dashed curve) and the photothermal effect (green dashed curve).

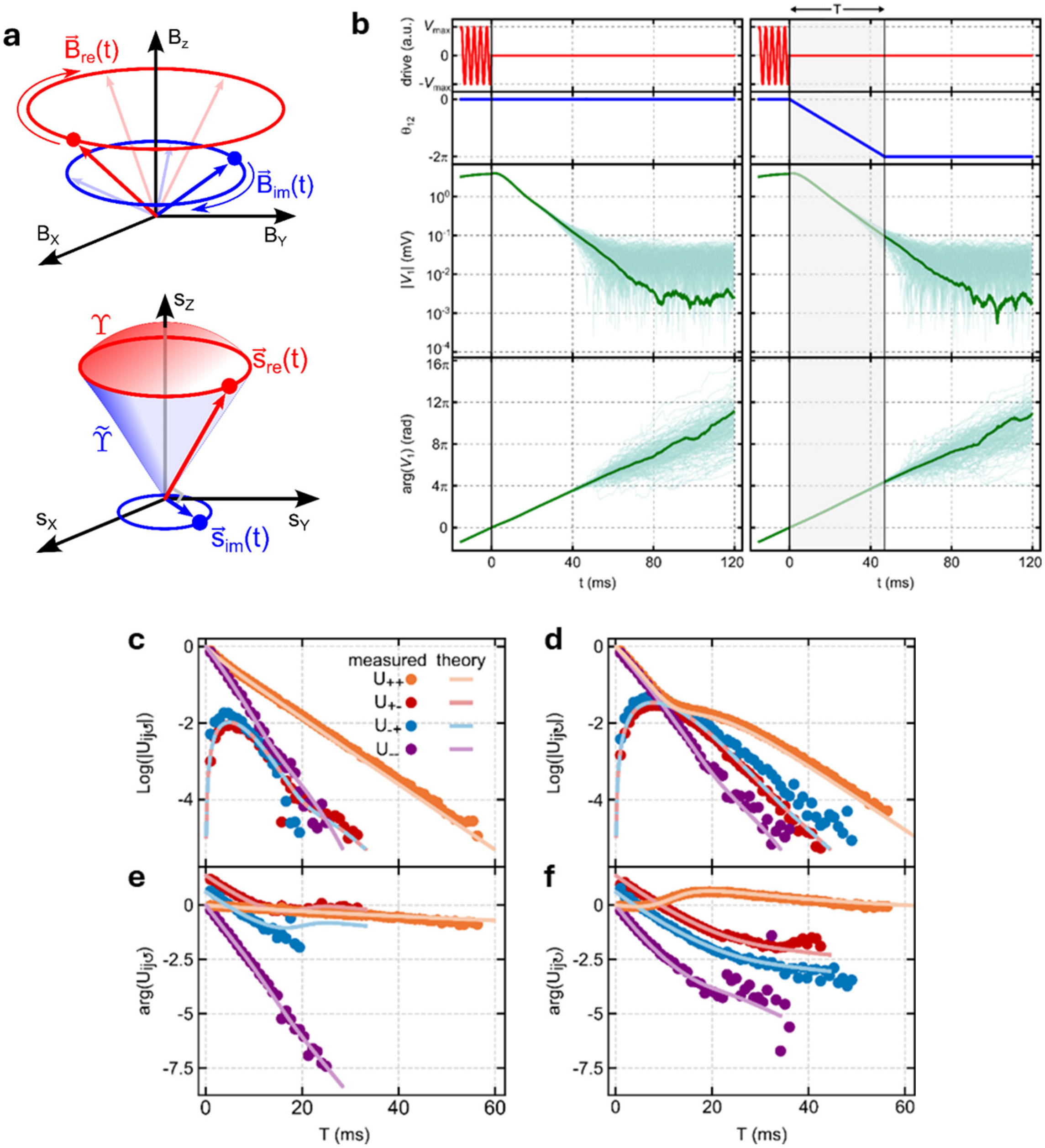

**Supplemental Figure 3: (a)** Two ways to visualize "simple" control loops. Left: in terms of real and imaginary magnetic fields, as in Fig. 2b. Right: in terms of the vectors $\vec{s}_{\mathrm{re}}$ and $\vec{s}_{\mathrm{im}}$. **(b)** Averaging multiple records. Left column: measurements with no control loop. Right column: measurements with the control loop. Red: the driving force, which is switched off at $t = 0$. Blue: $\theta_{12}(t)$. Green: magnitude and phase of the heterodyne signals $V_{1,2}(t)$. For each individual trace (faint green lines) the phase of the heterodyne signal is set to 0 for $t = 0$ so that the average of the traces (solid green line) reveals motion due to the drive. **(c-f)** Propagator elements. The amplitude (c,d) and phase (e,f) of $U(T)$ for the "simple" control loops with $P_1 = P_2 = 15\,\mu$W, $\delta/2\pi = -1$ MHz, $\eta = -50$ Hz. Points: data. Lines: no-free-parameters calculations. (c,e) show The "forward" version of the loop, defined by $\theta_{12}(t) = 2\pi t/T$. (d,f) show the "backward" version of the loop, defined by $\theta_{12}(t) = -2\pi t/T$. Data is shown only for $|U_{ii}(T)| > 5 \times 10^{-3}$.

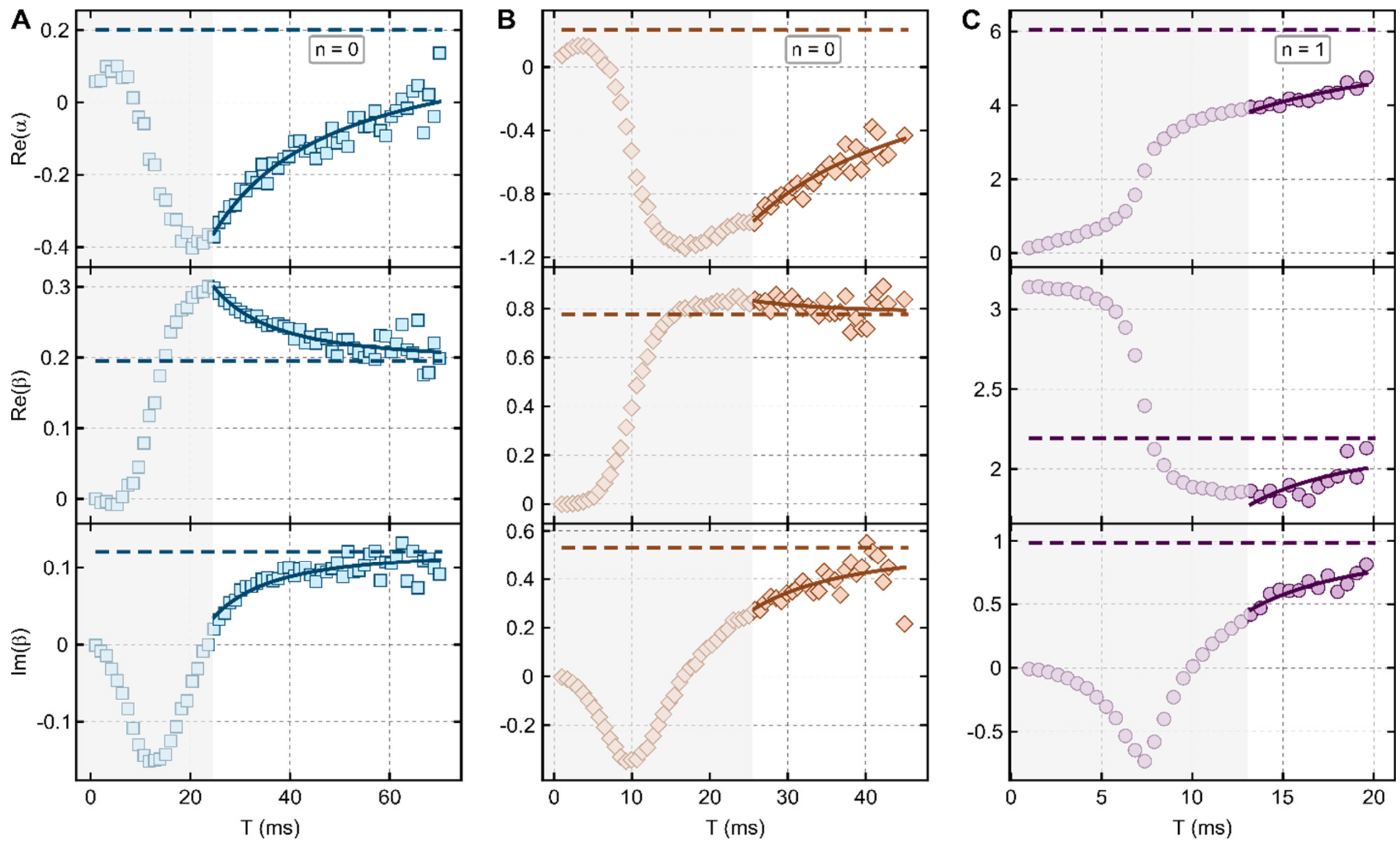

**Supplemental Figure 4:** Examples of fits to the asymptotic behavior of $\mathrm{Re}(\alpha(T))$ and $\beta(T)$, and the determination of $\phi_{\mathrm{B}}$. Shaded region: data excluded from the fit. **(a)** An example of a measurement in which 2/3 of the data was included in the fit. This is Option (3) of Supplemental Note 4. **(b)** An example of a measurement in which data were excluded for $T < 8/(|\lambda_{+} - \lambda_{-}|)$. This is Option (1) of Supplemental Note 4. **(c)** An example of a measurement in which 1/3 of the data was included in the fit. This is Option (2) of Supplemental Note 4. Top row: $\mathrm{Re}(\alpha(T))$, Middle and bottom rows: $\mathrm{Re}(\beta(T))$ and $\mathrm{Im}(\beta(T))$ respectively. Points: data, solid curve: fit, dashed line: fitted asymptote, which is used to calculate $\phi_{\mathrm{B}}$. For $\mathrm{Re}(\alpha(T))$, this asymptote is binned to determine the integer $n$ (Supplemental Note 4). The data in panels a, b, c were taken for control loops with $(P, \delta, \eta) = (17.5\ \mu\mathrm{W}, -1.5\ \mathrm{MHz}, -50\ \mathrm{Hz})$, $(17.5\ \mu\mathrm{W}, -0.75\ \mathrm{MHz}, -50\ \mathrm{Hz})$, $(20\ \mu\mathrm{W}, -0.25\ \mathrm{MHz}, -50\ \mathrm{Hz})$ respectively. All of these control loops used $\theta_{12}(t) = \pm 2\pi t/T$.

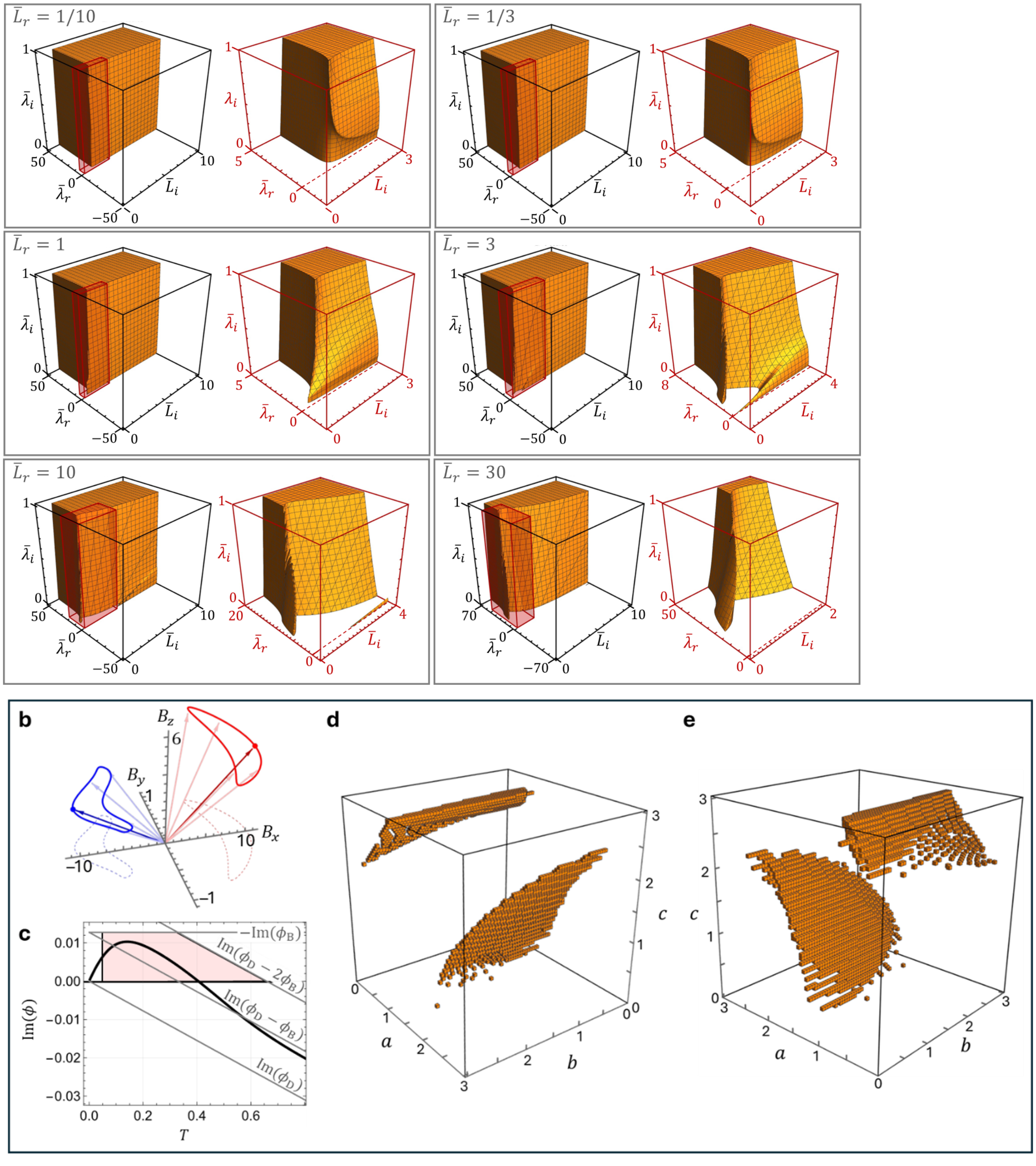

**Supplemental Figure 5: (a)** The range of "simple" loops that result in SSGG. Orange: the range of $\bar{\lambda}_\mathrm{r}, \bar{\lambda}_\mathrm{i}, \bar{L}_\mathrm{r}, \bar{L}_\mathrm{i}$ that can produce SSGG (the overbar indicates normalization by $|\mathcal{T}_\mathrm{i}|$). In each panel, the right plot shows a magnified view of the region shown in red in the left plot. **(b)** The non-simple loop parameterized by Eq. S79 with $(a, b, c) = (1.7, 0.4, 1.3)$. Red: $\mathrm{Re}[\vec{B}(s)]$. Blue: $\mathrm{Im}[\vec{B}(s)]$. Dashed curves: their projection on the $B_x$-$B_y$ plane. Circles: the start and stop of the loop. **(c)** Black curve: $\mathrm{Im}[\phi(T)]$ for $(a, b, c) = (0.6, 0.6, 2.7)$. Pink: the region $\Gamma$. $T_\mathrm{ad} = 0.047$ for these parameters. **(d,e)** Orange: the values of $(a, b, c)$ that produce SSGG. The two panels show different views of the same results.

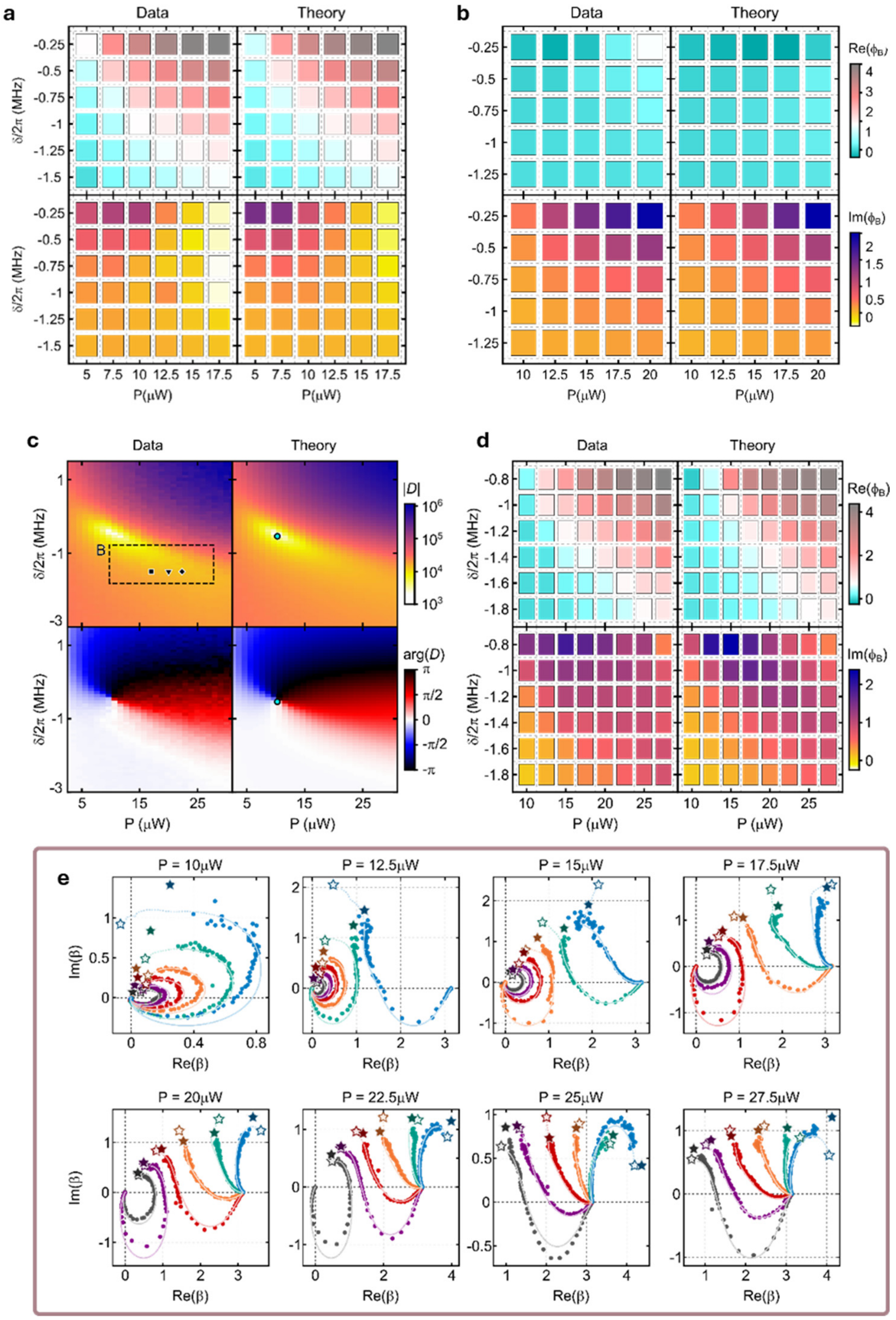

**Supplemental Figure 6: (a,b)** As in Fig. 3, but with $\eta/2\pi = -20$ Hz (a) and $\eta/2\pi = -90$ Hz (b). **(c-e)** Measurements using the membrane's (3,3) and (5,3) modes. **(c)** Absolute value and argument of the complex discriminant $D$ as a function of the laser power $P$ and detuning $\delta$ for $\eta/2\pi = -27.5$ Hz (similar to Fig. 1d). The square, triangle, and diamond show the parameters used for the measurements described in Supplemental Figure 10. **(d)** Left: $\phi_{\mathrm{B}}$ extracted from fits to $\beta(T)$ for control loops with $P$, $\delta$ drawn from the dashed box shown in (c). Right: predicted $\phi_{\mathrm{B}}$ (similar to Fig. 3c). **(e)** Parametric plots of $\beta(T)$ for all the measurements shown in d) (similar to Fig. 3a,b).

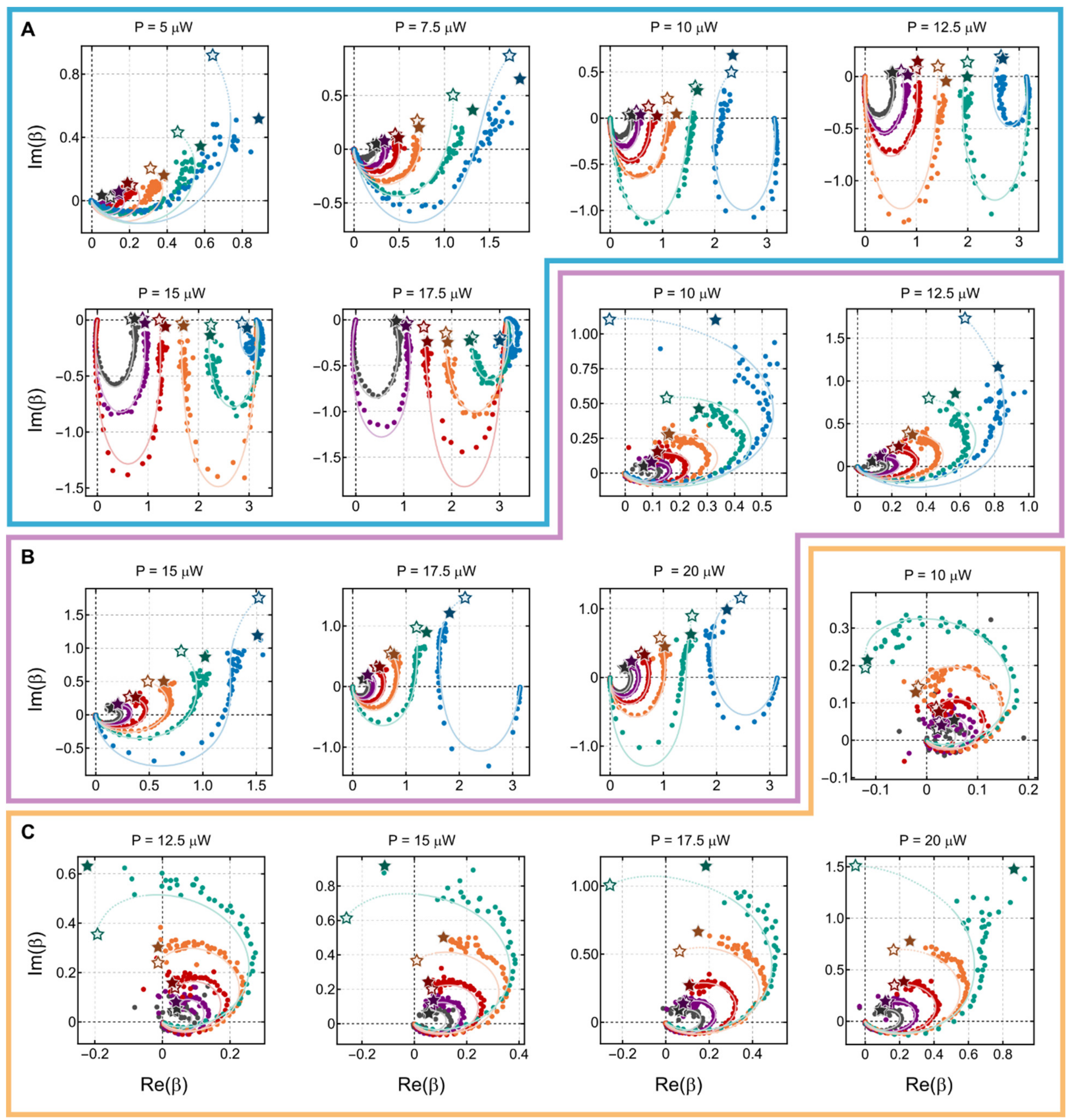

**Supplemental Figure 7:** Parametric plots of $\beta(T)$ for all "simple" loops, i.e. for those defined by $\theta_{12}(t) = \pm 2\pi t/T$. **(a)** The data corresponding to Supplemental Figure 6a (for which $\eta/2\pi = -20$ Hz). **(b)** The data corresponding to Fig. 3c (for which $\eta/2\pi = -50$ Hz). **(c)** The data corresponding to Supplemental Figure 6b (for which $\eta/2\pi = -90$ Hz). For every measurement for which the fit to $\mathrm{Re}(\alpha)$ returns $n = 1$, we have added $\pi$ to all data points in $\beta(T)$ as described in Supplemental Note 4.

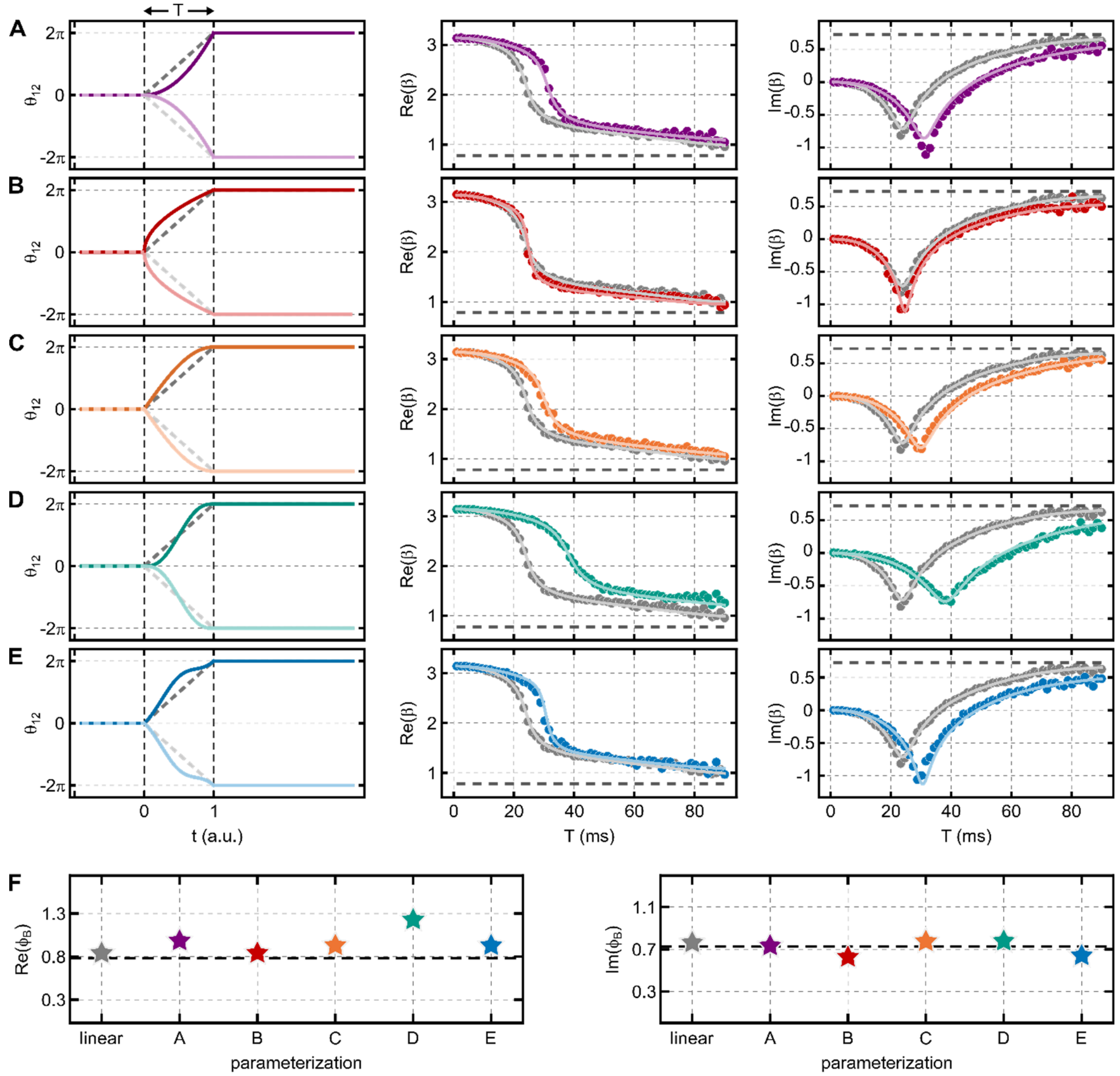

**Supplemental Figure 8:** Control loops with various $\theta_{12}(t)$. **(a-e)** Left column: the five $\theta_{12}(t)$ described in Supplemental Note 6 (solid). For comparison, the dashed line shows the "linear ramp" $\theta_{12}(t) = \pm 2\pi t/T$ (which is used for the measurements shown in Fig. 3). Center column: real part of $\beta(T)$ measured for the corresponding $\theta_{12}(t)$ (colored circles) along with no-free-parameters simulations (colored lines). The dashed line shows the predicted $\phi_{\mathrm{B}}$. For comparison, the results for the linear ramp are shown in gray (circles: data, curves: simulation). Right column: the same as the center column, but showing the imaginary part of $\beta(T)$. **(f)** The real (left) and imaginary (right) components of $\phi_{\mathrm{B}}$, extracted from asymptotic fits to $\beta(T)$ for each $\theta_{12}(t)$. The horizontal axis indicates the various $\theta_{12}(t)$. Dashed line: predicted value. These measurements used the membrane's (3,3) and (5,3) modes, and $(P_1, P_2, \Delta/2\pi, \eta/2\pi) =$ (21 μW, 21 μW, –1.6 MHz, –27.5 Hz).

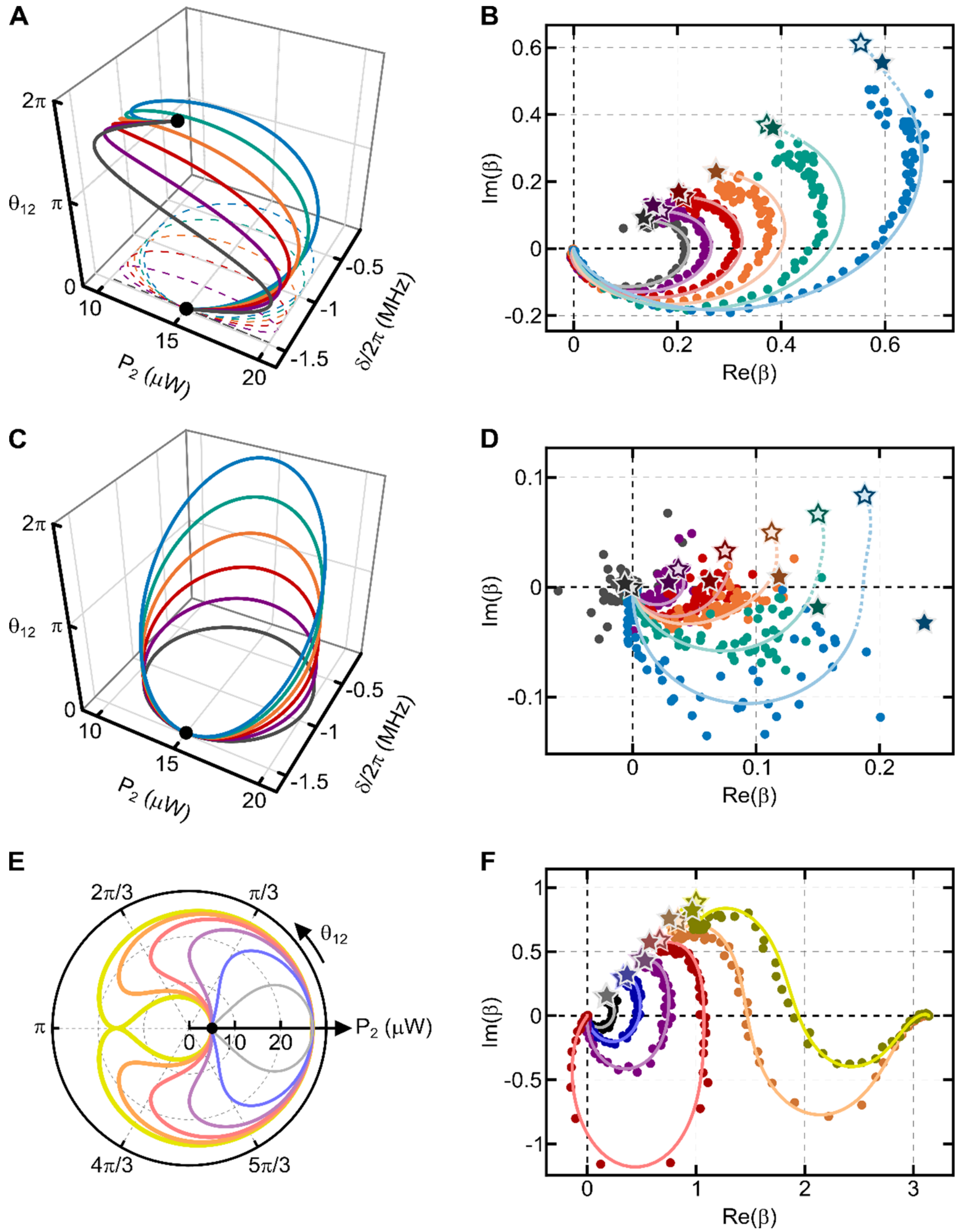

**Supplemental Figure 9:** Non-simple control loops. **(a)** Loops of the form $\vec{X}_1$, plotted in the 3D space of the parameters $\theta_{12}, P_2, \delta$. The solid lines are the actual loop, and the dashed lines show the loop projected onto the $(P_2, \delta)$ plane. Black dots are the start/stop point. **(b)** $\beta$ and $\phi_{\mathrm{B}}$ for the loops in (a). Circles: measurements of $\beta$. Curves: predicted $\beta$ (solid for the range of $T$ corresponding to the data; dashed for $T$ beyond the measured range). Solid stars: $\phi_{\mathrm{B}}$ extracted from fitting the $\beta(T)$ for large $T$. Open stars: predicted $\phi_{\mathrm{B}}$. The colors in (b) correspond to the loops in (a). **(c)** Loops of the form $\vec{X}_2$. **(d)** $\beta$ and $\phi_{\mathrm{B}}$ for the loops in (c). **(e)** Loops of the form $\vec{X}_3$. **(f)** $\beta$ and $\phi_{\mathrm{B}}$ for the loops in (e).

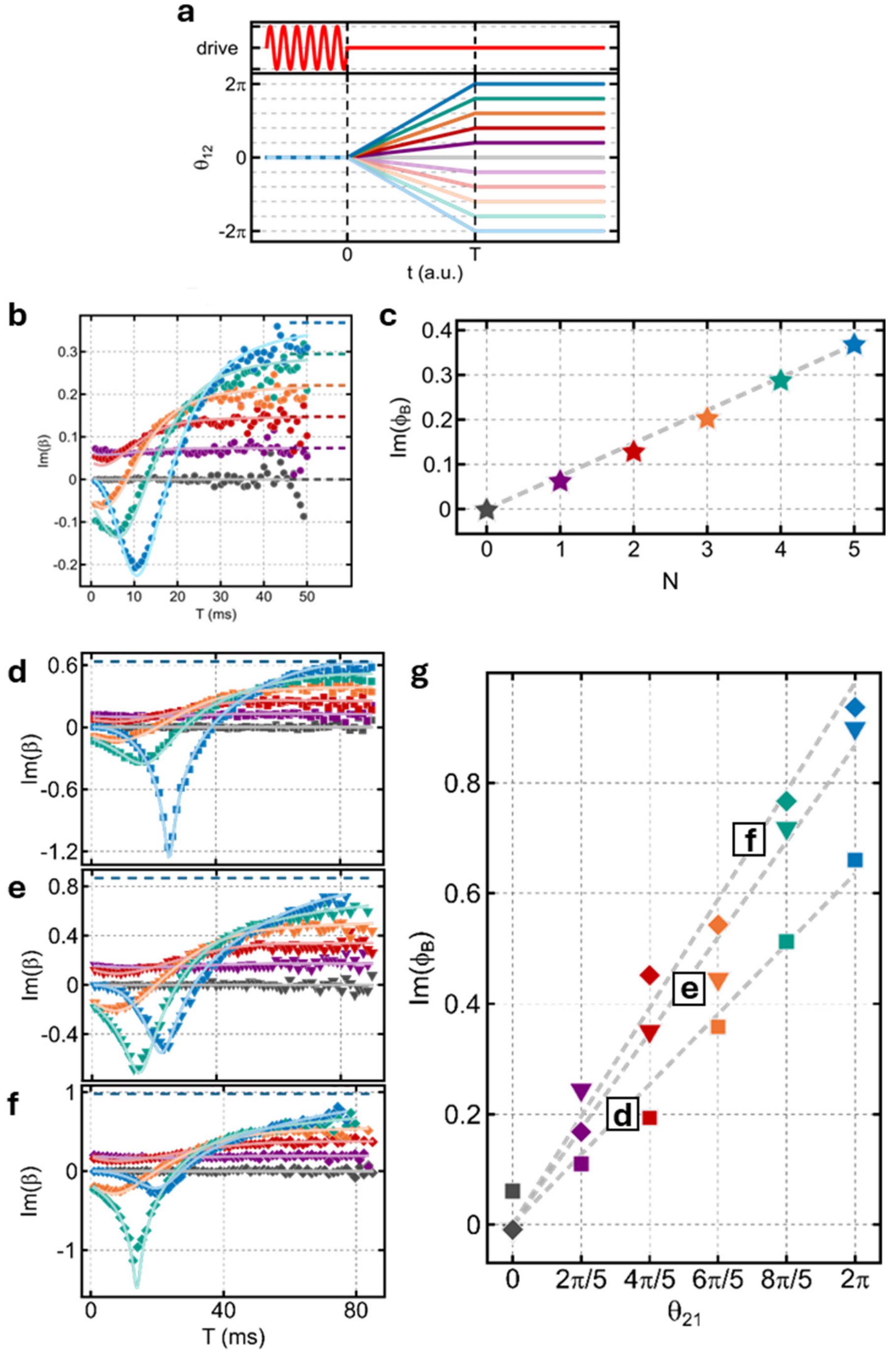

**Supplemental Figure 10: (a)** Control paths where $\theta_{12}$ runs from 0 to $\pm(N/5)2\pi$ for $N = 0,1,\ldots,5$. Solid (pale) lines: positive (negative) $\dot{\theta}_{12}$. (**b**) $\mathrm{Im}\big(\beta(T)\big)$ measured (circles) and predicted (curves) for each $N$. Here $P_{1,2} = P = 15\ \mu\mathrm{W}, \frac{\delta}{2\pi} = -0.87$ MHz and $\frac{\eta}{2\pi} = -50$ Hz. Dashed lines: predicted $\mathrm{Im}(\phi_{\mathrm{B}})$ for each $N$. (**c**) Stars: measured $\mathrm{Im}(\phi_{\mathrm{B}})$; Line: predicted $\mathrm{Im}(\phi_{\mathrm{B}})$. **(d-f)** as in (b) but with $\frac{\delta}{2\pi} = -1.5$ MHz, $\eta/2\pi = -27.5$ Hz, and $P = (17\ \mu\mathrm{W}, 20\ \mu\mathrm{W}, 23\ \mu\mathrm{W})$ for (d), (e), (f) respectively (these are the values marked in Supplemental Figure 6a). **(g)** $\mathrm{Im}(\phi_{\mathrm{B}})$ determined from the data in (d-f) versus $N$.

**Supplemental References:**